\documentclass[twocolumn,english,prl,twocolumn,floatfix,footinbib,superscriptaddress,showpacs, showkeys,notitlepage]{revtex4-1}
\usepackage[T1]{fontenc}
\usepackage[latin9]{inputenc}
\usepackage{float}
\usepackage{amsmath}
\usepackage{amssymb}
\usepackage{graphicx}
\usepackage{esint}

\makeatletter

\usepackage{babel}
 \usepackage[colorlinks]{hyperref}
\usepackage[explicit]{titlesec}

\usepackage{babel}

\makeatother

\usepackage{babel}
\begin{document}

\title{Steady-state properties of a thermodynamically unbalanced Fermi gas }

\author{Pedro Ribeiro}
\email{ribeiro.pedro@gmail.com}

\selectlanguage{english}%

\affiliation{CeFEMA, Instituto Superior Técnico, Universidade de Lisboa Av. Rovisco
Pais, 1049-001 Lisboa, Portugal}
\begin{abstract}
The current-carrying steady-state that arises in the middle of a metallic
wire connected to macroscopic leads is characterized regarding its
response functions, correlations and entanglement entropy. The spectral
function and the dynamical structure factor show clear non-equilibrium
signatures accessible by state-of-the-art techniques. In contrast
with the equilibrium case, the entanglement entropy is extensive with
logarithmic corrections at zero-temperature that depend on the wire-leads
coupling and, in a non-analytic way, on voltage. This shows that some
robust universal quantities found in gapless equilibrium phases do
not persist away from equilibrium.
\end{abstract}


\maketitle
\noindent\begin{minipage}[t]{1\columnwidth}%
\global\long\def\ket#1{\left| #1\right\rangle }

\global\long\def\bra#1{\left\langle #1 \right|}

\global\long\def\kket#1{\left\Vert #1\right\rangle }

\global\long\def\bbra#1{\left\langle #1\right\Vert }

\global\long\def\braket#1#2{\left\langle #1\right. \left| #2 \right\rangle }

\global\long\def\bbrakket#1#2{\left\langle #1\right. \left\Vert #2\right\rangle }

\global\long\def\av#1{\left\langle #1 \right\rangle }

\global\long\def\tr{\text{tr}}

\global\long\def\Tr{\text{Tr}}

\global\long\def\pd{\partial}

\global\long\def\im{\text{Im}}

\global\long\def\re{\text{Re}}

\global\long\def\sgn{\text{sgn}}

\global\long\def\Det{\text{Det}}

\global\long\def\abs#1{\left|#1\right|}

\global\long\def\up{\uparrow}

\global\long\def\down{\downarrow}

\global\long\def\k{\mathbf{k}}

\global\long\def\wks{\mathbf{\omega k}\sigma}

\global\long\def\vc#1{\mathbf{#1}}

\global\long\def\bs#1{\boldsymbol{#1}}

\global\long\def\t#1{\text{#1}}
\end{minipage}Current-carrying steady-states (CCSS) are characterized by a steady
flow of equilibrium-conserved quantities, such as energy, spin or
charge. Of direct relevance to transport experiments are steady currents
generated by coupling a system to reservoirs at different thermodynamic
potentials. The resulting CCSS are thermodynamically unbalanced, i.e.
do not fulfill equilibrium fluctuation dissipation relations \cite{Ribeiro2013,Ribeiro2015b}.
CCSS in one or quasi-one-dimensional systems are of relevance in many
fields, including charge and spin transport in electronic devices
and in cold atom setups. 

Due to kinetic constraints, accounting for relaxation in one dimension
requires to go beyond 2-body interaction terms and explicitly account
for 3- and higher-body collisions \cite{Levchenko2011,Micklitz2011,Micklitz2012}
and thus may be neglected for weakly interacting clean samples. For
non-interacting electrons on a wire, ideal reservoirs can be mimicked
by injecting particles from plus and minus infinity with a given energy
distributions\cite{Datta1995,Imry1997,diVentra2008,Kamenev2011}.
This ideal conditions, alluded to as Landauer reservoirs \cite{Landauer1957,Buttiker1986},
yield to a local energy distribution function that is the average
of those of the leads. A series of studies featuring non-equilibrium
Luttinger liquids \cite{Gutman2008,Gutman2009,Dinh2010,Takei2010,Gutman2011,Popkov2012}
found that interaction-induced dephasing may smear the local energy
distribution even in the absence of relaxation. In the presence of
a strong enough relaxation the system is expected to equilibrate locally.
Treatments based on the Boltzaman equation have been used to obtain
the distribution function of the charge carriers in this regime \cite{Lunde2007,Rech2009,Lunde2009b,Levchenko2011,Micklitz2011,Micklitz2012}. 

Experiments featuring CCSS, designed to access the local energy distribution
of charge caries, were performed using tunneling spectroscopy in mesoscopic
wires \cite{Pothier1997a,Pothier1997b,Anthore2003} and carbon nanotubes
\cite{Chen2009}. The local energy distribution, measured in the center
of the wire, was reported to exhibit a characteristic double step
form resulting from contribution of both Fermi functions of the electronic
leads. The sharp steps seen at low temperatures are smeared out as
temperature increases or in the presence of electron-electron interactions,
disorder or electron-phonon coupling.

The study of current-carrying states recently became available for
cold atomic setups \cite{Brantut2012}. Mainly motivated by these
advances, a rather different body of works investigated the time evolution
of two initially disconnected semi-infinite wires held at different
equilibrium conditions. After the two wires are connected a CCSS forms
around the connection point. This central region grows with time,
with the remaining parts of the wire acting essentially as reservoirs.
At large times a translational invariant CCSS is locally observed
\cite{Lancaster2010,Lancaster2011,Sabetta2013,Bernard2012,Bernard2014,Calabrese2008,Eisler2013,Alba2014,Ogata2002,Karevski2009}.
Interestingly, some of the properties of the CCSS created in this
way, in particular the momentum-resolved electronic distribution,
are similar to the CCSS obtained using a Landauer description \cite{Calabrese2008,Lancaster2010,Bernard2012,Eisler2013,Bernard2014,Alba2014,Viti2015}.
Furthermore, in both cases, the entanglement entropy of a region in
the middle or the wire yields to the same universal result as in equilibrium. 

Recently, momentum-resolved spectroscopic measurements became available
for non-equilibrium electronic systems \cite{Gierz2013,Cilento2014}.
Since to the same local energy spectrum may correspond various momentum
distributions, this developoments allows a better characterization
of the state and may shed some light on discrepancies between existing
theoretical predictions and experimental findings. In addition, transport
experiments in cold atomic setups \cite{Brantut2012} allow to access
a set of physical quantities that are difficult to study in solid-state
devices. These developments urge for a better theoretical understanding
of thermodynamically unbalanced CCSS, beyond the local energy distribution
function, that is currently still unavailable. 

This work addresses the CCSS realized on a finite metallic wire coupled
to metallic leads at different temperatures and chemical potentials.
The lead-wire couplings are treated explicitly as they induce additional
reflections that change the energy distribution of the carries \cite{Datta1995,Imry1997}.
At equilibrium, the system can be described by an one dimensional
electron-gas as we assume no electron-electron interactions or disorder
to be present. Our approach describes the low energy sector where
the dispersion relation is essentially linear and the reservoirs'
chemical potentials and temperatures are much smaller that the wire's
bandwidth. We study the one- and two- point functions and analyze
the entanglement content in the wire's central region.

\begin{figure}
\centering{}\includegraphics[width=1\columnwidth]{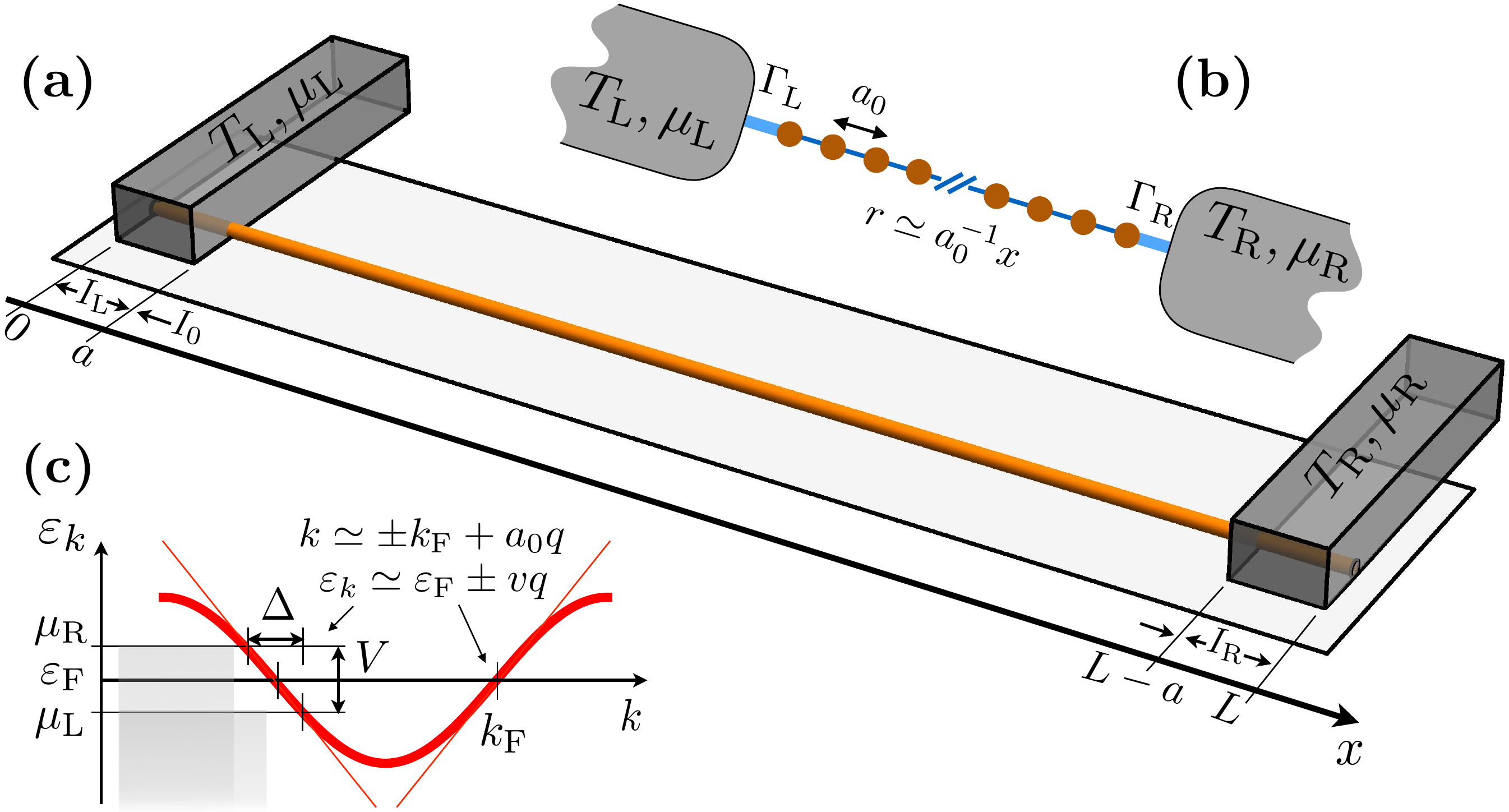}\caption{\label{fig:model-1}(a) Schematics of the setup. (b) Tight-binding
chain coupled to reservoirs. (c) Dispersion relation of the tight-binding
chain, $\varepsilon_{k}$, linearized around the Fermi-momentum, $k_{\text{F}}$,
defined such that $\varepsilon_{k_{\text{F}}}=\left(\mu_{\text{L}}+\mu_{\text{R}}\right)/2$,
with $V=\mu_{\text{L}}-\mu_{\text{R}}$ and $\Delta=a_{0}V/v$. }
\end{figure}

\textit{Model and Results.} - Consider the setup of a Fig.\ref{fig:model-1}-(a)
depicting a 1d fermionic gas on a wire of length $L$ attached to
external leads. In the wire, we assume fermions to have an approximately
linear dispersion, with velocity $v$, within a window of size $2\Lambda$
around the Fermi points. The effective Hamiltonian of the isolated
metallic wire, valid for energies scales below $v\Lambda$, is given
by 
\begin{eqnarray}
H & = & -iv\int dx\,\Psi^{\dagger}\left(x\right)\bs{\sigma}_{z}\pd_{x}\Psi\left(x\right),
\end{eqnarray}
where $\Psi\left(x\right)=\left\{ \psi_{\text{L}}\left(x\right),\psi_{\text{R}}\left(x\right)\right\} ^{T}$,
with $\psi_{l=\text{L},\text{R}}\left(x\right)$ corresponding to
the left and right moving fermions with $\left\{ \psi_{l}\left(x\right),\psi_{l'}^{\dagger}\left(x'\right)\right\} =\delta_{ll'}\delta\left(x-x'\right)$.
At position $x_{\text{L}}=0$ and $x_{\text{R}}=L$ the boundary conditions
are given by $\psi_{\text{L}}\left(x_{l}\right)=e^{i\phi_{l}}\psi_{\text{R}}\left(x_{l}\right)$
with phase shift $\phi_{l}$. To model the leads we assume that the
extremities of the wire are connected to fermionic reservoirs within
a region of length $a\ll L$ such as in Fig.\ref{fig:model-1}-(a).
The reservoirs are assumed to be metallic, with a bandwidth much larger
than any characteristic energy scale of the wire. Their chemical potentials
$\mu_{l=\t L,\t R}$ and temperatures $T_{l=\t L,\t R}=\beta_{l=\t L,\t R}^{-1}$
are taken to be much smaller than the energy cutoff $v\Lambda$. The
hybridization of the wire with reservoir $l$ is characterized by
the energy scale $av\gamma_{l}$, see \cite{SupMat}. The corresponding
timescale $\left(av\gamma_{l}\right)^{-1}$ gives the characteristic
time for a particle in region $I_{l=\t L,\t R}$ to escape the reservoir
$l$. In the following we set $\mu_{\text{L}}-\mu_{\text{R}}=V\ge0$
without lost of generality.

\begin{figure}
\centering{}\includegraphics[width=1\columnwidth]{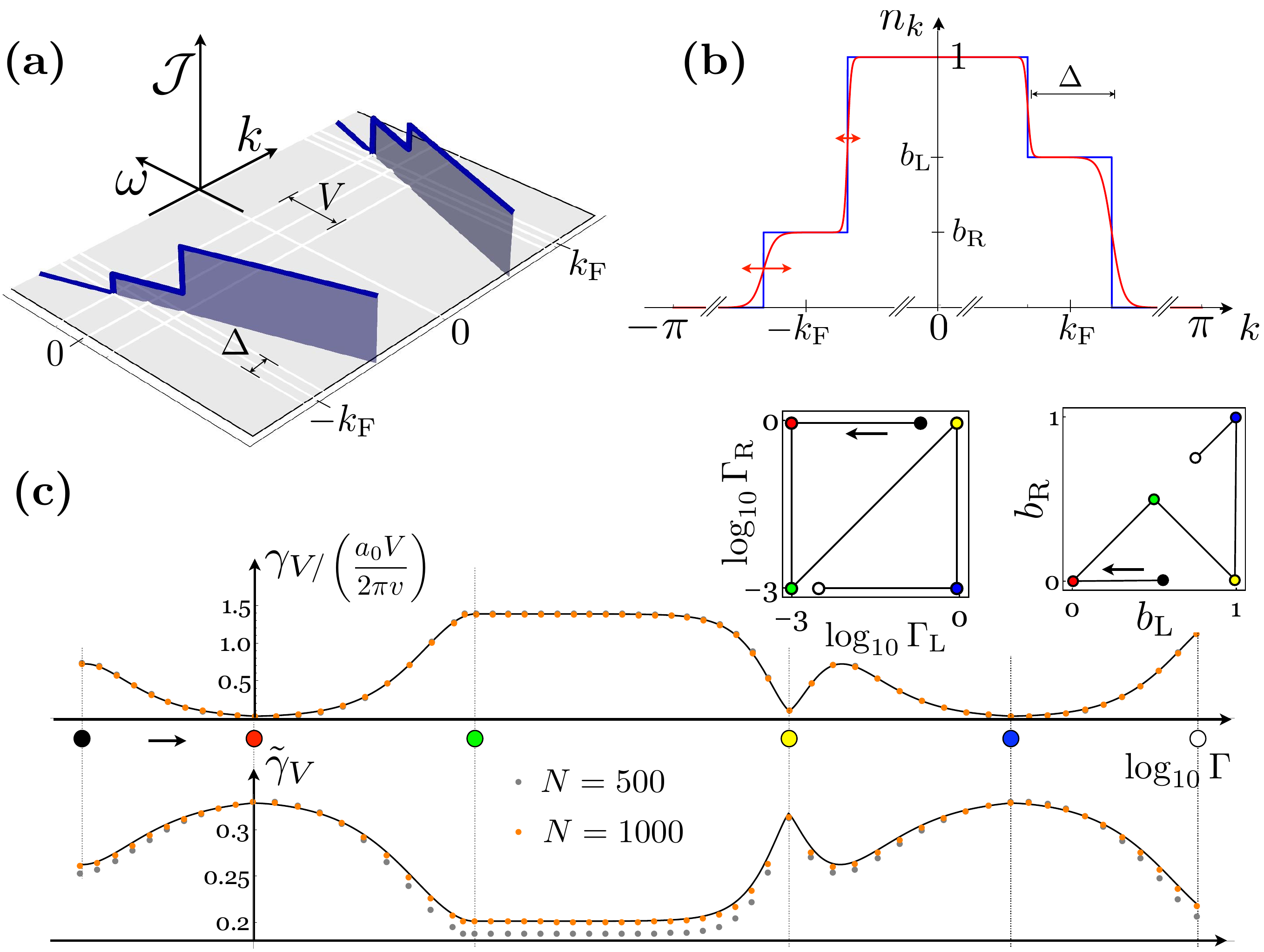}\caption{\label{fig:1_particle}(a) $\mathcal{J}_{k}\left(\omega\right)=-i\frac{1}{2\pi}G_{k}^{<}\left(\omega\right)$,
proportional to the transition rate measured by angle-resolved photo
emission, for $T_{\text{L}},T_{\text{R}}=0$. (b) Momentum resolved
occupation number $n_{k}$ given for $T_{\text{L}}=T_{\text{R}}=0$
(blue) and $T_{\text{L}}>T_{\text{R}}\protect\neq0$ (red). (c) Extensive
coefficient of the entanglement entropy $\gamma_{\text{V}}$ and the
logarithmic correction $\tilde{\gamma}_{\text{V}}$ computed for set
of points $\{\Gamma_{\text{L}},\Gamma_{\text{R}}\}$ given in the
inset following the color code. The second inset depicts the corresponding
$\{b_{\text{L}},b_{\text{R}}\}$ values. The numerical values obtained
for the tight-binding model with $\varepsilon_{\protect\t F}=.3t$,
$V=.2t$, and $T_{\text{L}}=T_{\text{R}}=0$ are compared with the
low energy theory predictions (black line). }
\end{figure}

To help validate our analytic treatment we present a set of numerical
results for a tight-binding model on a chain of $N$ sites, with Hamiltonian
$H_{\text{TB}}=-t\sum_{r=1}^{N}c_{r}^{\dagger}c_{r+1}+\text{h.c.}$,
coupled at the two end sites to a wide-band reservoir, as in Fig.\ref{fig:1_particle}-(b).
The reservoir $l$ introduces an hybridization energy scale $\Gamma_{l}=\pi t_{l}^{2}D_{l}$,
where $t_{l}$ is the chain-reservoir hopping and $D_{l}$ is the
local density of states of the reservoir \cite{Ribeiro2015a,SupMat}.
The average Fermi momentum $k_{\text{F}}$ is defined such that $\varepsilon_{k_{\text{F}}}=\varepsilon_{\text{F}}=\left(\mu_{\text{L}}+\mu_{\text{R}}\right)/2$,
thus for $q$ smaller than $\Lambda$: $\varepsilon_{k}\simeq\varepsilon_{\text{F}}\pm vq$
and $c_{r}\simeq e^{ik_{\text{F}}r}\psi_{\text{L}}\left(a_{0}r\right)+e^{-ik_{\text{F}}r}\psi_{\text{R}}\left(a_{0}r\right)$,
with $k=\pm k_{\text{F}}+a_{0}q$ and $a_{0}$ the lattice constant,
see Fig.\ref{fig:1_particle}-(c). Further identification between
the continuum and tight-binding models yield to: $N=La_{0}^{-1}$,
$k_{\text{F}}=\arccos\left(-\varepsilon_{\text{F}}/2t\right)$, $v=2a_{0}t\sin k_{\text{F}}$.
and $\phi_{L}=2k_{\text{F}}-\pi$, $\phi_{R}=-2k_{\text{F}}La_{0}^{-1}-\pi$.
The relation $a\gamma_{l}=\frac{1}{4}\ln\left(-\frac{\Gamma_{l}^{2}t^{-2}-2t^{-1}\sin k_{\t F}\Gamma_{l}+1}{\Gamma_{l}^{2}t^{-2}+2t^{-1}\sin k_{\t F}\Gamma_{l}+1}\right)$
between the hybridization constants can be derived by matching the
imaginary part of the wave vectors \cite{SupMat}.

In order to address the properties of the steady-state that forms
under the conditions described above we compute the retarded ($R$),
advanced ($A$) and Keldysh ($K$) components of the Green's function
in the frequency domain. The wide-band nature of the reservoirs considerably
simplifies our treatment \cite{Ribeiro2015a}. In this case, the retarded
Green's function is given by $\bs G^{R}\left(\omega;x,x'\right)=\bra x\left(\omega-\bs K\right)^{-1}\ket{x'}$,
where $\bs K=\int dx\,\ket x-iv\left[\bs{\sigma}_{z}\pd_{x}+\sum_{l=\text{L},\text{R}}\gamma_{l}\Theta\left[\abs{x-x_{l}}-a\right]\right]\bra x$
is a non-hermitian operator describing a single particle on a wire
with particle sinks in regions $I_{l=\t R,\t L}$. $\bs K$ is diagonalized
by left and right eigenvectors $\bra{\tilde{\psi}_{n}}$ and $\ket{\psi_{n}}$,
with eigenvalues $\lambda_{n}=vq_{n}$. In position space $\braket{\tilde{\psi}_{n}}x$
and $\braket x{\psi_{n}}$ are plane waves, with a complex valued
momentum $q_{n}$, for $x$ within the regions $I_{l=R,0,L}$ defined
in Fig.\ref{fig:model-1}-(a). The wave amplitudes within each region
and the quantization condition $q_{n}\equiv-\frac{1}{2L}\left(\phi_{\text{L}}-\phi_{\text{R}}\right)+\frac{\pi n}{L}-ia\frac{\gamma_{\text{L}}+\gamma_{\text{R}}}{L}$,
with $n\in\mathbb{Z}$, are determined by imposing boundary conditions
at $x=0,L$, the continuity of the wave functions at $x=a,L-a$ and
the normalization of the wave-functions $\braket{\tilde{\psi}_{n}}{\psi_{n'}}=\delta_{nn'}$.
The Keldysh component of the Green's function in the steady state,
given by $\bs G^{K}=\bs G^{R}\bs{\Sigma}^{K}\bs G^{A}$, can also
be obtained explicitly using essentially the same procedure \cite{SupMat}.
For convenience, in the following we analyze the hermitian matrices
$\bs{\rho}^{-}=-\left[\bs G^{R}-\bs G^{A}\right]/2\pi i$ and $\bs{\rho}^{+}=\bs G^{K}/\left(-2\pi i\right)$
rather than the Green's functions. 

We concentrate in the middle region of the wire in the limit $L\to\infty$,
in which case the quantities $\bs{\rho}_{\text{bulk}}^{\pm}\left(\omega;x,x'\right)\equiv\lim_{L\to\infty}\bs{\rho}^{\pm}\left(\omega;x+L/2,x'+L/2\right)$
become translational invariant, for finite $x$ and $x'$, and their
Fourier components are given by 
\begin{eqnarray}
\bs{\rho}_{\text{bulk }}^{\pm}\left(\omega,q\right) & = & \text{diag}\left\{ \rho_{\text{L}}^{\pm}\left(\omega,q\right),\rho_{\text{R}}^{\pm}\left(\omega,q\right)\right\} \label{eq:rho_bulk}
\end{eqnarray}
with $\rho_{\text{L}/\text{R}}^{-}\left(\omega,q\right)=\delta\left(\omega\mp vq\right)$
and $\rho_{l}^{+}\left(\omega,q\right)=\left[1-2n_{l}\left(\omega\right)\right]\rho_{l}^{-}\left(\omega,q\right)$,
for $\abs q<\Lambda$, and where
\begin{eqnarray}
n_{l}\left(\omega\right) & = & \frac{b_{l}}{e^{\beta_{l}\left(\omega-\mu_{l}\right)}+1}+\frac{\left(1-b_{l}\right)}{e^{\beta_{\bar{l}}\left(\omega-\mu_{\bar{l}}\right)}+1}
\end{eqnarray}
is the energy distribution function of the $l$-movers, with $\bar{R}=L$
, $\bar{L}=R$, $a\gamma_{\text{L}}=\frac{1}{4}\ln\frac{1-\mathit{b}_{\text{R}}}{1-\mathit{b}_{\text{L}}}$
and $a\gamma_{\text{R}}=\frac{1}{4}\ln\frac{\mathit{b}_{\text{L}}}{\mathit{b}_{\text{R}}}$.
Since $V\ge0$ we have that $\mathit{b}_{\text{L}}\ge\mathit{b}_{\text{R}}$.
In \cite{SupMat} we provide additional results for $x$ and $x'$
near the extremities of the wire, where the state is not translational
invariant in the $L\to\infty$ limit. Using the correspondence with
the continuum model, the tight-binding Green's functions $G_{rr'}^{\alpha}\left(\omega\right)$,
or equivalently the quantities $\rho_{rr'}^{\pm}\left(\omega\right)$,
are given by $\rho_{rr'}^{\pm}\left(\omega\right)=\int_{-\pi}^{\pi}\frac{dk}{2\pi}\rho_{k}^{\pm}\left(\omega\right)e^{ik\left(r-r'\right)},$
for $r$ and $r'$ in the middle of the wire, where 

\begin{multline}
\rho_{k}^{\pm}\left(\omega\right)=\rho_{\text{L }}^{\pm}\left[\omega,\left(k-k_{\text{F}}\right)a_{0}^{-1}\right]\Theta\left(\abs{k-k_{\text{F}}}-\Lambda a_{0}\right)\\
+\rho_{\text{R }}^{\pm}\left(\omega,k+k_{\text{F}}\right)\Theta\left(\abs{k+k_{\text{F}}}-\Lambda a_{0}\right).\label{eq:rho_k}
\end{multline}
Due to the quadratic nature of the model, these quantities can be
used to compute all correlations and response functions restricted
to the center of the wire and to low energies. In particular, the
single-particle density matrix $\varrho_{rr'}\left(t\right)=\av{c_{r}^{\dagger}\left(t\right)c_{r'}\left(t\right)}$,
that in the steady state is given by $\varrho_{rr'}=-\pi\int\frac{d\omega}{2\pi}\rho_{r'r}^{+}\left(\omega\right)+\frac{1}{2}\delta_{rr'}$,
can be approximated by 
\begin{eqnarray}
\varrho_{rr'} & \simeq\varrho_{r-r'}= & \int_{-\pi}^{\pi}\frac{dk}{2\pi}e^{-i\left(r-r'\right)k}n_{k},
\end{eqnarray}
with 
\begin{multline}
n_{k}=n_{\text{L}}\left[\frac{v}{a_{0}}\left(k-k_{F}\right)\right]\Theta\left(\abs{k-k_{\text{F}}}-\Lambda a_{0}\right)\\
+n_{\text{R}}\left[\frac{v}{a_{0}}\left(-k_{F}-k\right)\right]\Theta\left(\abs{k+k_{\text{F}}}-\Lambda a_{0}\right),
\end{multline}
the occupation number of momentum $k$. This relations can be complemented
by $n_{k}=0$ or $n_{k}=1$ away from the range of validity of the
low energy theory, i.e. $\abs{k\pm k_{\text{F}}}>\Lambda a_{0}$.
Fig. \ref{fig:1_particle}-(b) shows $n_{k}$ for $T_{\text{L}},T_{\text{R}}=0$
and for finite but distinct $T_{\text{L}}$ and $T_{\text{R}}$. For
finite $V$ there is a double step structure around each Fermi point
with width $\Delta=a_{0}V/v$ and height $b_{\t L}$ ($b_{\t R}$)
for $k$ near $k_{\text{F}}$ ($-k_{\text{F}}$). A single step per
Fermi point is recovered in three different cases: for reflection-less
leads, with $b_{\t L}=1,b_{\t R}=0$, reproducing the results obtained
using Landauer reservoirs; when one of the leads effectively decouples,
$b_{\t L}=1,b_{\t R}=1$ ($b_{\t L}=0,b_{\t R}=0$), i.e. the wire
coupling to the left (right) lead is much larger than that of the
right (left), in which case the wire distribution function becomes
that of the lead which strongly couples to the system. The steps are
smoothen with the temperature associated to the respective of the
reservoirs. For $T_{\text{L}},T_{\text{R}}\neq0$, $\varrho_{r}\propto e^{-\abs r/\xi}$
decays exponentially in $r$, with a temperature-dependent characteristic
length $\xi$ that diverges as $T_{\text{L}}$ or $T_{\text{R}}$
vanish. For $T_{\text{L}},T_{\text{R}}=0$, one obtains $\varrho_{r}=\frac{1}{r}\left\{ \sin\left(\frac{\Delta r}{2}\right)\right.\left[\left(b_{\text{L}}-1\right)e^{-irk_{\text{F}}}+b_{\text{R}}e^{irk_{\text{F}}}\right]+e^{-\frac{1}{2}i\Delta r}\left.\sin\left(rk_{\text{F}}\right)\right\} $,
corresponding to a $1/r$ decay as in the equilibrium $T=0$ case.
For reflection-less leads, i.e. $b_{\text{L}}=1$, $b_{\text{R}}=0$
, the argument of $\varrho_{r}$ becomes linear in $r$: $\arg\left(\varrho_{r}\right)=-\frac{1}{2}\Delta r$
as observed in the CCSS formed after a quench in the Refs. \cite{Lancaster2010,Sabetta2013}.
The local distribution function, as measured by tunneling spectroscopy,
is $n_{\text{local}}\left(\omega\right)=\left[n_{\text{R}}\left(\omega\right)+n_{\text{L}}\left(\omega\right)\right]/2$,
which for equal contacts (i.e. $b_{\t L}+b_{\t R}=1$) becomes independent
on $b_{l}$. Tunneling spectroscopy measurements performed with similar
contacts can therefore be insensitive to some of the features of $n_{k}$.
The particle current $J=-it\av{c_{r}^{\dagger}c_{r+1}-c_{r+1}^{\dagger}c_{r}}$,
given in the low energy sector by $J\simeq\frac{1}{2\pi}\left(b_{\text{L}}-b_{\text{R}}\right)V$,
is independent of $v$ and of the temperature, assuming for consistency
$\Delta\ll1$. 

In addition to static quantities the CCSS is characterized by its
dynamic correlators. Fig. \ref{fig:1_particle}-(a) depicts the one
point function $\mathcal{J}_{k}\left(\omega\right)=-i\frac{1}{2\pi}G_{k}^{<}\left(\omega\right)=\frac{1}{2}\left[\rho_{k}^{-}\left(\omega\right)-\rho_{k}^{+}\left(\omega\right)\right]$,
that is proportional to the transition rate as measured by angle-resolved
photo emission. For $\abs{\omega}<\Lambda v$ it can be approximated
by $\mathcal{J}_{k}\left(\omega\right)\simeq n_{k}\left\{ \delta\left[\omega-va_{0}^{-1}\left(k-k_{\text{F}}\right)\right]\right.+\left.\delta\left[\omega+va_{0}^{-1}\left(k+k_{\text{F}}\right)\right]\right\} $,
thus the step structure of Fig. \ref{fig:1_particle}-(a) is the same
of $n_{k}$.  Even at zero temperature, there is a non-vanishing
probability of finding a particle above $\varepsilon_{\t F}$ for
$V>0$ and thus $\mathcal{J}_{k}\left(\omega>0\right)$ does not vanish. 

We now turn to the entanglement properties of the CCSS. The entanglement
entropy of a region $\Sigma$ is defined as $S_{\Sigma}=-\tr\left(\hat{\rho}_{\Sigma}\ln\hat{\rho}_{\Sigma}\right)$
where $\hat{\rho}_{\Sigma}=\tr_{\bar{\Sigma}}\hat{\rho}$ is obtained
from the total density matrix $\hat{\rho}$ by trancing out the degrees
of freedom belonging to $\bar{\Sigma}$, the complement of $\Sigma$.
In the following we consider a region $\Sigma_{\ell}$ of size $\ell$
in the center of the wire and we denote $S_{\ell}=S_{\Sigma_{\ell}}$.
Since the model is Gaussian $S_{\ell}=\tr\left[s\left(\varrho_{\ell}\right)\right]$
, with $s\left(\nu\right)=-\nu\ln\nu-\left(1-\nu\right)\ln\left(1-\nu\right)$,
can be computed from the single particle density matrix $\varrho_{\ell}=\sum_{rr'\in\Sigma_{\ell}}\ket r\varrho_{rr'}\bra r$
restricted to $\Sigma_{\ell}$ . $S_{\ell}$ can be calculated, following
Ref. \cite{Its2009}, using asymptotic results for approximating determinants
of Toplitz matrices, valid in the $\ell\to\infty$ limit. For $T_{R},T_{L}\neq0$
we find, employing Szegö's limit theorem, $S_{\ell}=-\ell\sum_{l}\int_{-\infty}^{\infty}\frac{dk}{2\pi}\,s\left[n_{k}\right]+\gamma_{\text{A}}$,
where $\gamma_{\text{A}}$ is an $\ell$ independent constant. For
the case $T_{R},T_{L}=0$ we have to appeal to the Fisher-Hartwing
conjecture (see \cite{Its2009}) for the case of Fig.\ref{fig:1_particle}-(b)-(blue
line) where $n_{k}$ has four discontinuities rather then the two
present at equilibrium. Flowing the same steps as in Ref. \cite{Its2009}
we obtain, (see \cite{SupMat}), $S_{\ell}\simeq\gamma_{\text{V}}\ell+\tilde{\gamma}_{\text{V}}\ln\left(\ell\right)+\gamma_{\text{A}}+O\left(1/\ell\right)$,
where $\gamma_{\text{V}}=\frac{a_{0}V}{2\pi v}\left[s\left(b_{\text{L}}\right)+s\left(b_{\text{R}}\right)\right]$
is the coefficient of the volume term, and $\tilde{\gamma}_{\text{V}}=\frac{1}{3}-\tilde{s}\left(b_{L}\right)-\tilde{s}\left(b_{R}\right)$
with
\begin{multline*}
\tilde{s}\left(b\right)=\frac{1}{24}+\frac{1}{4\pi^{2}}\left[\left(2\mathit{b}-1\right)\left[\text{Li}_{2}\left(1-\mathit{b}\right)-\text{Li}_{2}\left(\mathit{b}\right)\right]\right.\\
\left.+\left(1-\mathit{b}\right)\log^{2}\left(1-\mathit{b}\right)+\mathit{b}\log^{2}\left(\mathit{b}\right)+\log\left(\mathit{b}\right)\log\left(1-\mathit{b}\right)\right],
\end{multline*}
is the coefficient of the logarithmic correction that is voltage-independent.
The mutual information $S\left(A,B\right)=S_{A}+S_{B}-S_{A+B}$ between
two adjacent segments of length $\ell/2$, is given by $S\left(A,B\right)=\tilde{\gamma}_{\text{V}}\ln\left(\ell\right)+\gamma_{\text{A}}-2\tilde{\gamma}_{\text{V}}\ln\left(2\right)+O\left(1/\ell\right)$.
Here the volume term cancels and we can easily access $\tilde{\gamma}_{\text{V}}$
numerically. Fig.\ref{fig:1_particle}-(c) shows $\gamma_{\text{V}}$
and $\tilde{\gamma}_{\text{V}}$ as a function of the hybridization
along a path in the $\Gamma_{\t L}-\Gamma_{\t R}$ plane as well as
the corresponding values in the $b_{\t L}-b_{\t R}$ plane. With increasing
system size the numerical results obtained for the tight-binding model
converge to the analytic predictions. $\gamma_{\text{V}}$ and $\tilde{\gamma}_{\text{V}}$
vary in the opposite sense - a large volume term corresponds to a
small mutual information content. The maximum value $\tilde{\gamma}_{\text{V}}=1/3$,
obtained at equilibrium, is attained for $V>0$ whenever $b_{\t L}$
and $b_{\t R}$ are such that only two discontinuities arise in $n_{k}$,
in which case $\gamma_{\text{V}}=0$. For a generic point on the $b_{\t L}-b_{\t R}$
plane with $\tilde{\gamma}_{\text{V}}\neq1/3$ the limit $V\to0$
is singular since for $V=0$, and only at that point, one recovers
the equilibrium value $\tilde{\gamma}_{\text{V}}=1/3$ (numerics shown
in \cite{SupMat}). This singularity prohibits the calculation of
the entanglement entropy perturbatively in $V$ and thus this quantity
cannot be obtained from linear response arguments. 
\begin{figure}
\centering{}\includegraphics[width=1\columnwidth]{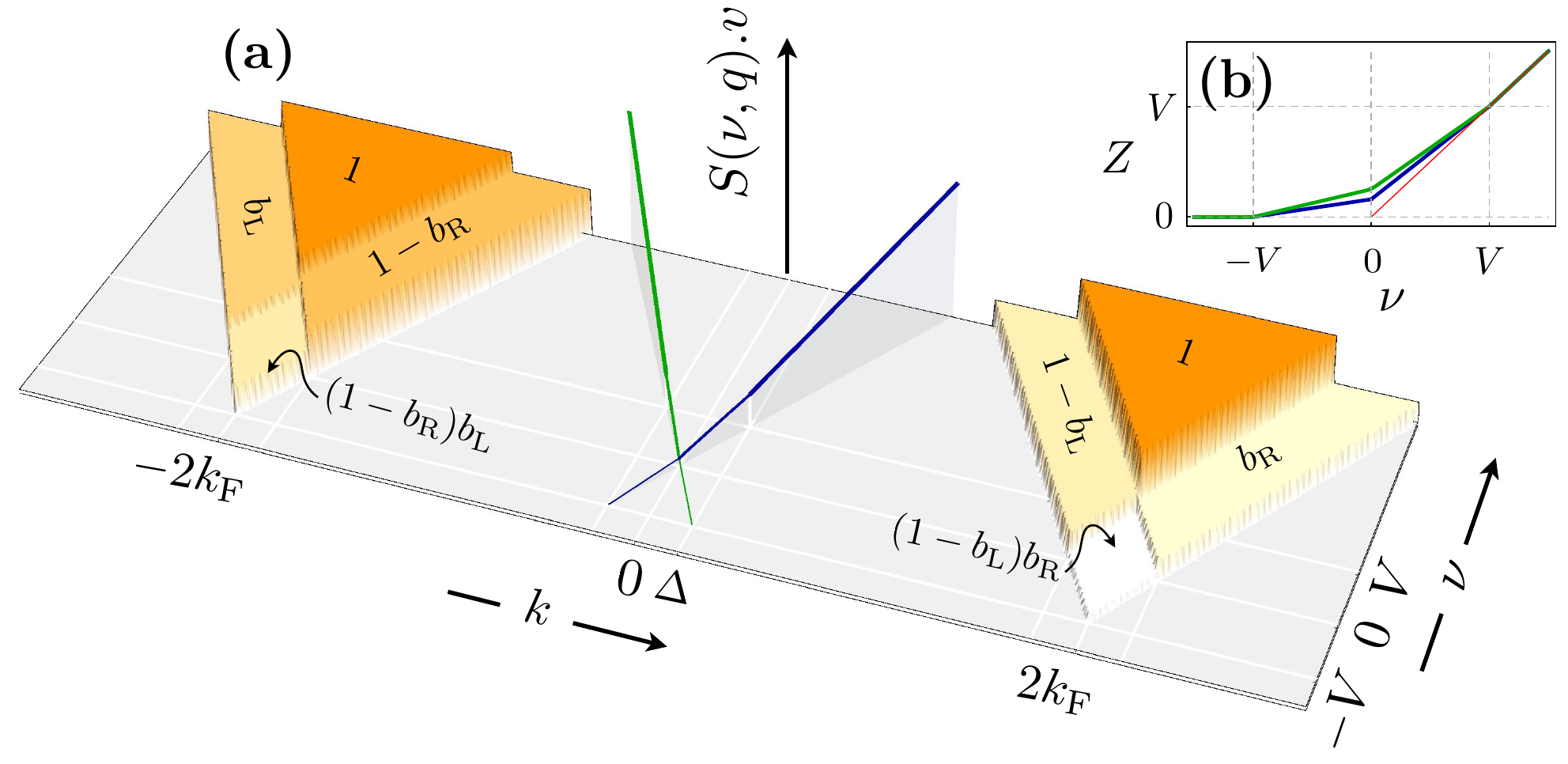}\caption{\label{fig:2_particle} (a) Dynamical structure factor $S_{p}\left(\nu\right)=i\chi_{p}^{>}\left(\nu\right)$,
directly accessible by neutron scattering. (b) Wight of the $\delta$-function
$Z$ for the positive (blue) and negative (green) velocity branches
near $k=0$. The red line depicts the equilibrium, i.e. $V=0$, result. }
\end{figure}

Finally we analyze the two-point density-density correlations and
response functions of the CCSS encoded in the lesser and grater components
of the charge susceptibility $\chi_{rr'}^{>}\left(t,t'\right)=-i\av{\hat{n}_{r}(t)\hat{n}_{r'}(t')}$,
$\chi_{rr'}^{<}\left(t,t'\right)=-i\av{\hat{n}_{r'}(t')\hat{n}_{r}(t)}$.
As for the one point function, for $r,r'$ in the central region of
the wire, $\chi_{rr'}^{><}$ become approximately translational invariant.
In this regime, the Fourier transformed quantities $\chi_{p}^{\pm}\left(\nu\right)=-\frac{1}{2\pi i}\left[\chi_{p}^{>}\left(\nu\right)\pm\chi_{p}^{<}\left(\nu\right)\right]$
are given by
\begin{multline}
\chi_{p}^{\pm}\left(\nu\right)=\left[\chi_{\text{LL}}^{\pm}\left(\nu,a_{0}^{-1}p\right)+\chi_{\text{RR}}^{\pm}\left(\nu,a_{0}^{-1}p\right)\right]\\
+\chi_{\text{LR}}^{\pm}\left[\nu,a_{0}^{-1}\left(p-2k_{F}\right)\right]+\chi_{\text{RL}}^{\pm}\left[\nu,a_{0}^{-1}\left(p+2k_{F}\right)\right],
\end{multline}
where the first two terms correspond to low momentum ($\abs p<a_{0}\Lambda$)
and the last two terms to $2k_{\t F}$ contributions ($\abs{p-2k_{\text{F}}}<a_{0}\Lambda$
and $\abs{p-2k_{\text{F}}}<a_{0}\Lambda$) and where
\begin{eqnarray*}
\chi_{ll}^{+}\left(\nu q\right) & = & \frac{1}{2v}\delta\left(\nu\mp vq\right)\int\frac{d\omega}{2\pi}\left[1-F_{l}\left(\omega\right)F_{l}\left(\omega-\nu\right)\right],\\
\chi_{l\bar{l}}^{+}\left(\nu q\right) & = & \frac{1}{4\pi v}\left[1-F_{l}\left(\frac{\pm vq+\nu}{2}\right)F_{\bar{l}}\left(\frac{\pm vq-\nu}{2}\right)\right],\\
\chi_{ll}^{-}\left(\nu q\right) & = & \frac{1}{2v}\delta\left(\nu\mp vq\right)\int\frac{d\omega}{2\pi}\left[F_{l}\left(\omega\right)-F_{l}\left(\omega-\nu\right)\right],\\
\chi_{l\bar{l}}^{-}\left(\nu q\right) & = & \frac{1}{4\pi v}\left[F_{l}\left(\frac{\pm vq+\nu}{2}\right)-F_{\bar{l}}\left(\frac{\pm vq-\nu}{2}\right)\right],
\end{eqnarray*}
where the upper (lower) signs are for $l=\text{R}$ ($l=\t L$) and
$F_{l}\left(\omega\right)=1-2n_{l}\left(\omega\right)$. The dynamical
structure factor $S_{p}\left(\nu\right)=i\chi_{p}^{>}\left(\nu\right)$,
that can be directly accessed by neutron scattering is shown is Fig.
\ref{fig:2_particle} for $T_{\t L}=T_{\t R}=0$. As in equilibrium,
the contribution at low momentum is coherent, but the $\delta$-function
weight $Z$ acquires a nontrivial dependence on frequency as seen
in the inset. Contributions near $\pm k_{\t F}$, form the particle-hole
excitation continuum, also develop a step like structure dependent
on $b_{\t L}$ and $b_{\t R}$ that gets smeared out at finite temperatures.
In the reflection-less case $b_{\t L}=1,b_{\t R}=0$ the particle-hole
continuum is simply shift up (down) in energy for positive (negative)
momentum. 

\textit{Conclusion.} - The properties of a CCSS in a thermodynamically
unbalanced one-dimensional system are found to crucially depend on
the coupling to the leads thought the double-step form of the momentum
occupation number near each Fermi point. This feature imprints a clear
signature to the one- and two-point functions that are accessible
by state-of-the-art angle-resolved photo emission and neutron scattering
and can be used to characterize the non-equilibrium state. For a generic
CCSS, the entanglement entropy is found to be extensive and proportional
to the applied voltage. At zero temperature, the logarithmic pre-factor
of the mutual information of two adjacent segments shows that their
mutual entanglement is never larger than at equilibrium, being singular
at $V=0$ and voltage-independent for $\abs V>0$. Therefore, the
equilibrium result, that can be obtain by conformal filed theory arguments
for gapless one-dimensional phases with unit central charge, does
not hold away from equilibrium. This suggest that the physics away
from equilibrium may be ruled by low energy fixed point theories fundamentally
different from the equilibrium ones. 
\begin{acknowledgments}
We gratefully acknowledge discussions with D. Esteve, M. Haque, S.
Kirchner, A. Lazarides, A. Rubtsov, P. Sacramento and V. R. Vieira.
PR acknowledges support by FCT through the Investigador FCT contract
IF/00347/2014.
\end{acknowledgments}

\bibliographystyle{apsrev4-1}
\bibliography{../article/char_ness}

\appendix
\cleardoublepage

\setcounter{equation}{0}
\setcounter{section}{0}
\renewcommand{\theequation}{SM--\arabic{equation}}
\titleformat{\section}{\normalfont}{\thesection}{1em}{\textbf{\stepcounter{section}\setcounter{subsection}{0} \arabic{section} -- \MakeUppercase{#1}}}
\titleformat{\subsection}{\normalfont}{\thesubsection}{1em}{\textbf{ \stepcounter{subsection}\setcounter{subsubsection}{0} \arabic{section}.\arabic{subsection} -- #1} }

\clearpage{}

\begin{widetext}

\begin{center}
\Large \bf {-- Supplemental Material -- \\ Steady-state properties of a thermodynamically unbalanced Fermi gas }\bigskip
\par\end{center}

\begin{center}
Pedro Ribeiro$^1$\medskip\it \small \\
$^1$CeFEMA, Instituto Superior Técnico, Universidade de Lisboa Av. Rovisco Pais, 1049-001 Lisboa, Portugal
\par\end{center}

In the following supplemental material we provide additional details
of the analytical and numerical analysis performed in the main text.
In particular we provide: a derivation of the relations between quantities
of the tight-binding and of the continuum model; the derivation of
the retarded, advanced and Keldysh Green's functions, and, of the
infinite volume limit; we also provide additional the numerical results,
for finite and infinite volume, of the one-point function; we present
the detailed derivation of the entanglement entropy and give some
additional numerical results; and we derive of the two-point function.

\setcounter{equation}{0}
\setcounter{section}{0}
\renewcommand{\theequation}{SM--\arabic{equation}}
\titleformat{\section}{\normalfont}{\thesection}{1em}{\textbf{\stepcounter{section}\setcounter{subsection}{0} \arabic{section} -- \MakeUppercase{#1}}}
\titleformat{\subsection}{\normalfont}{\thesubsection}{1em}{\textbf{ \stepcounter{subsection}\setcounter{subsubsection}{0} \arabic{section}.\arabic{subsection} -- #1} }

\section{Identifications with the Tight-Binding model I - Green's Functions}

\subsection{Green's functions}

Using that, in the low energy sector, $c_{r}\simeq e^{ik_{\text{F}}r}\psi_{\text{L}}\left(a_{0}r\right)+e^{-ik_{\text{F}}r}\psi_{\text{R}}\left(a_{0}r\right)$
the Green's function of the tight-binding model on the Keldysh contour
can be approximated by 

\begin{eqnarray}
G_{rr'}\left(z,z'\right) & \equiv & -i\av{T_{\gamma}c_{r}\left(z\right).c_{r'}^{\dagger}\left(z'\right)}\simeq\left(\begin{array}{cc}
e^{ik_{\text{F}}r} & e^{-ik_{\text{F}}r}\end{array}\right).\bs G\left(zra_{0}^{-1},z'r'a_{0}^{-1}\right).\left(\begin{array}{c}
e^{-ik_{\text{F}}r'}\\
e^{ik_{\text{F}}r'}
\end{array}\right),\label{eq:id_1}
\end{eqnarray}
where 

\begin{eqnarray}
\bs G\left(zx,z'x'\right) & \equiv & -i\av{T_{\gamma}\Psi\left(zx\right).\Psi^{\dagger}\left(z'x'\right)},
\end{eqnarray}
is the continuum Green's function and $T_{\gamma}$ is the time ordered
operator for two times $z$ and $z'$ on the Keldysh contour $\gamma$.
We use the standard definitions of larger and lesser Green's functions,
$\bs G^{>(<)}\left(tx,t'x'\right)\equiv\bs G\left(tx,t'x'\right)$
for $z=t$ ($z'=t'$) coming after $z'=t'$ ($z=t$) along $\gamma$,
and 
\begin{eqnarray}
\mathbf{G}^{R}(tx,t'x') & \equiv & \Theta(t-t')\left[\mathbf{G}^{>}(tx,t'x')-\mathbf{G}^{<}(tx,t'x')\right],\\
\mathbf{G}^{A}(tx,t'x') & \equiv & -\Theta(t'-t)\left[\mathbf{G}^{>}(tx,t'x')-\mathbf{G}^{<}(tx,t'x')\right],\\
\mathbf{G}^{K}(tx,t'x') & \equiv & \mathbf{G}^{>}(tx,t'x')+\mathbf{G}^{<}(tx,t'x').
\end{eqnarray}
Below we will use the notation $\mathbf{G}^{a}(tx,t'x')=\bra x\mathbf{G}^{a}(t,t')\ket{x'}$,
for $a=R,A,K$, and in the steady-state we define $\bs G^{a}\left(\omega\right)=\int\frac{d\omega}{2\pi}e^{i\omega\left(t-t'\right)}\bs G^{a}\left(t,t'\right)$
. For convenience we work with the quantities 

\begin{eqnarray}
\bs{\rho}^{-}\left(\omega\right) & = & -\left[\bs G^{R}\left(\omega\right)-\bs G^{A}\left(\omega\right)\right]/\left(2\pi i\right),\\
\bs{\rho}^{+}\left(\omega\right) & = & -\bs G^{K}\left(\omega\right)/\left(2\pi i\right),
\end{eqnarray}
rather than the Green's functions themselves. These are proportional
to the spectral function and to the imaginary part of the Keldysh
Green's function, respectively, and encode the same physical information.
Note that defined in this way both $\rho^{\pm}$ are hermitian matrices
$\left(\rho^{\pm}\right)^{\dagger}=\rho^{\pm}$. As in Eq.(\ref{eq:id_1}),
we also have the relation 
\begin{eqnarray}
\rho_{rr'}^{\pm}\left(\omega\right) & \simeq & \left(\begin{array}{cc}
e^{ik_{\text{F}}r} & e^{-ik_{\text{F}}r}\end{array}\right).\bs{\rho}^{\pm}\left(\omega;ra_{0}^{-1},r'a_{0}^{-1}\right).\left(\begin{array}{c}
e^{-ik_{\text{F}}r'}\\
e^{ik_{\text{F}}r'}
\end{array}\right),
\end{eqnarray}
between continuum and tight-binding quantities. 

\section{Single particle correlation functions}

\subsection{Self-energies }

The reservoirs are assumed to be metallic with a bandwidth much larger
than any characteristic energy scale of the wire. In this limit the
retarded ($R$) and advanced ($A$) components of the systems's self-energy
due to the reservoir $l$ are given by (see \cite{Ribeiro2015a})
\[
\bs{\Sigma}_{l}^{R/A}\left(\omega;xx'\right)=\mp i\delta\left(x-x'\right)v\,\gamma_{l}\Theta\left[\abs{x-x_{l}}-a\right]
\]
 where $\Theta\left(x\right)$ is the Heaviside theta-function, $\gamma_{l}$
is a constant that characterizes the hybridization of the wire with
reservoir $l$. Since the reservoirs are taken to be at thermal equilibrium
the Keldysh ($K$) component is given by 
\begin{eqnarray*}
\bs{\Sigma}_{l}^{K}\left(\omega;xx'\right) & = & \tanh\left[\frac{\beta_{l}}{2}\left(\omega-\mu_{l}\right)\right]\left[\Sigma_{l}^{R}\left(\omega;xx'\right)-\Sigma_{l}^{A}\left(\omega;xx'\right)\right].
\end{eqnarray*}
The total self-energy is the sum of the contributions of both reservoirs
$\bs{\Sigma}^{R/A/K}=\sum_{l}\bs{\Sigma}_{l}^{R/A/K}$. 

\subsection{Diagonalization of the $K$ operator }

Following Ref.\cite{Ribeiro2015a} we can write the retarded steady-state
Green's function in the wide band approximation as $G^{R}\left(\omega\right)=\left(\omega-K\right)^{-1}$,
where the operator $\bs K$ is given by 
\begin{eqnarray}
\bs K & = & \int dx\,\ket x\left[-iv\sigma_{z}\pd_{x}-iv\,\sigma_{0}\sum_{l}\gamma_{l}\Theta\left(\abs{x-x_{l}}-a\right)\right]\bra x,
\end{eqnarray}
with boundary conditions imposed by $\bs S_{l}\braket{x_{l}}{\psi}=\braket{x_{l}}{\psi}$
with
\begin{eqnarray}
\bs S_{l} & = & \left(\begin{array}{cc}
0 & e^{i\phi_{l}}\\
e^{-i\phi_{l}} & 0
\end{array}\right).
\end{eqnarray}
For $\ket{\psi}$ and $\bra{\tilde{\psi}}$, respectively right and
left eigenvectors $\bs K$, the relations $\bra x\bs K\ket{\psi}=\lambda\braket x{\psi}$
and $\bra{\tilde{\psi}}\bs K\ket x=\lambda\braket{\tilde{\psi}}x$
imply that, for $x$ within the regions $I_{l=R,0,L}$, defined in
Fig.1-(a) in the main text,
\begin{eqnarray}
\braket{x\in I_{l}}{\psi} & = & \left(\begin{array}{c}
A_{l}e^{i\left(q+i\gamma_{l}\right)x}\\
B_{l}e^{-i\left(q+i\gamma_{l}\right)x}
\end{array}\right);\\
\braket{\tilde{\psi}}{x\in I_{l}} & = & \left(\begin{array}{cc}
\tilde{A}_{l}e^{-i\left(q+i\gamma_{l}\right)x} & \tilde{B}_{l}e^{i\left(q+i\gamma_{l}\right)x}\end{array}\right);
\end{eqnarray}
with $\lambda=vq$ and $\gamma_{0}=0$. From the boundary conditions
at $x_{\text{L}}=0$ and $x_{\text{R}}=L$ we have 
\begin{eqnarray}
B_{\text{L}} & = & e^{-i\phi_{\text{L}}}A_{\text{L}};\\
B_{\text{R}} & = & e^{-i\phi_{\text{R}}}e^{2i\left(q+i\gamma_{\text{R}}\right)L}A_{\text{R}};\\
\tilde{B}_{\text{L}} & = & e^{i\phi_{\text{L}}}\tilde{A}_{\text{L}};\\
\tilde{B}_{\text{R}} & = & e^{i\phi_{\text{R}}}e^{-2i\left(q+i\gamma_{\text{R}}\right)L}\tilde{A}_{\text{R}}.
\end{eqnarray}
In the same way, by ensuring the continuity of the wave function at
$x=a$ and $x=L-a$ we obtain
\begin{eqnarray}
\left(\begin{array}{c}
A_{\text{L}}e^{i\left(q+i\gamma_{\text{L}}\right)a}\\
B_{\text{L}}e^{-i\left(q+i\gamma_{\text{L}}\right)a}
\end{array}\right) & = & \left(\begin{array}{c}
A_{0}e^{iqa}\\
B_{0}e^{-iqa}
\end{array}\right);\\
\left(\begin{array}{c}
A_{0}e^{iq\left(L-a\right)}\\
B_{0}e^{-iq\left(L-a\right)}
\end{array}\right) & = & \left(\begin{array}{c}
A_{\text{R}}e^{i\left(q+i\gamma_{\text{R}}\right)\left(L-a\right)}\\
B_{\text{R}}e^{-i\left(q+i\gamma_{\text{R}}\right)\left(L-a\right)}
\end{array}\right);\\
\left(\begin{array}{cc}
\tilde{A}_{\text{L}}e^{-i\left(q+i\gamma_{\text{L}}\right)a} & \tilde{B}_{\text{L}}e^{i\left(q+i\gamma_{\text{L}}\right)a}\end{array}\right) & = & \left(\begin{array}{cc}
\tilde{A}_{0}e^{-iqa} & \tilde{B}_{0}e^{iqa}\end{array}\right);\\
\left(\begin{array}{cc}
\tilde{A}_{0}e^{-iq\left(L-a\right)} & \tilde{B_{0}}e^{iq\left(L-a\right)}\end{array}\right) & = & \left(\begin{array}{cc}
\tilde{A}_{\text{R}}e^{-i\left(q+i\gamma_{\text{R}}\right)\left(L-a\right)} & \tilde{B}_{\text{R}}e^{i\left(q+i\gamma_{\text{R}}\right)\left(L-a\right)}\end{array}\right);
\end{eqnarray}
Combining these results yields to 
\begin{eqnarray}
A_{0} & = & B_{0}e^{i\phi_{\text{L}}}e^{-2\gamma_{\text{L}}a};\\
A_{0} & = & B_{0}e^{-2iqL}e^{i\phi_{\text{R}}}e^{2\gamma_{\text{R}}a};\\
\tilde{A}_{0} & = & \tilde{B}_{0}e^{-i\phi_{\text{L}}}e^{2\gamma_{\text{L}}a};\\
\tilde{A}_{0} & = & \tilde{B}_{0}e^{-i\phi_{\text{R}}}e^{2iqL}e^{-2\gamma_{\text{R}}a}.
\end{eqnarray}
Solving for the amplitudes of the central region one obtains 
\begin{eqnarray}
\left(\begin{array}{cc}
1 & e^{i\phi_{\text{L}}}e^{-2\gamma_{\text{L}}a}\\
e^{2iqL}e^{-i\phi_{\text{R}}}e^{-2\gamma_{\text{R}}a} & 1
\end{array}\right)\left(\begin{array}{c}
A_{0}\\
B_{0}
\end{array}\right) & = & 0
\end{eqnarray}
corresponding to the quantization condition 
\begin{eqnarray}
1-e^{i\left(\phi_{\text{L}}-\phi_{\text{R}}\right)}e^{-2\left(\gamma_{\text{L}}a+\gamma_{\text{R}}a\right)}e^{2iqL} & = & 0,
\end{eqnarray}
i.e.
\begin{eqnarray}
q_{n} & = & -\frac{1}{2L}\left(\phi_{\text{L}}-\phi_{\text{R}}\right)+\frac{\pi n}{L}-ia\frac{\gamma_{\text{L}}+\gamma_{\text{R}}}{L},\label{eq:quantization_cond}
\end{eqnarray}
with $n\in\mathbb{Z}$. Finally the normalization condition 
\begin{eqnarray}
\braket{\tilde{\psi}_{n}}{\psi_{n}} & = & \left(\tilde{A}_{0}A_{0}+\tilde{B}_{0}B_{0}\right)L=2\tilde{B}_{0}B_{0}L=1
\end{eqnarray}
 yields to
\begin{eqnarray}
\braket{x\in I_{L}}{\psi} & = & \frac{1}{\sqrt{2L}}\left(\begin{array}{c}
e^{(a-x)\gamma_{L}+iqx}\\
e^{(a+x)\gamma_{L}-i\left(\phi_{L}+qx\right)}
\end{array}\right)\\
\braket{x\in I_{0}}{\psi} & = & \frac{1}{\sqrt{2L}}\left(\begin{array}{c}
e^{iqx}\\
e^{-i\left(2ia\gamma_{L}+\phi_{L}+qx\right)}
\end{array}\right)\\
\braket{x\in I_{R}}{\psi} & = & \frac{1}{\sqrt{2L}}\left(\begin{array}{c}
e^{-\gamma_{R}(a-L+x)+iqx}\\
e^{-\gamma_{R}(a+L-x)+i\left(2Lq-qx-\phi_{R}\right)}
\end{array}\right)
\end{eqnarray}
and 
\begin{eqnarray}
\braket{\tilde{\psi}}{x\in I_{L}} & = & \frac{1}{\sqrt{2L}}\left(\begin{array}{cc}
e^{(x-a)\gamma_{L}-iqx} & e^{i\left(i(a+x)\gamma_{L}+\phi_{L}+qx\right)}\end{array}\right)\\
\braket{\tilde{\psi}}{x\in I_{0}} & = & \frac{1}{\sqrt{2L}}\left(\begin{array}{cc}
e^{-iqx} & e^{i\left(2ia\gamma_{L}+\phi_{L}+qx\right)}\end{array}\right)\\
\braket{\tilde{\psi}}{x\in I_{R}} & = & \frac{1}{\sqrt{2L}}\left(\begin{array}{cc}
e^{\gamma_{R}(a-L+x)-iqx} & e^{\gamma_{R}(a+L-x)-2iLq+iqx+i\phi_{R}}\end{array}\right)
\end{eqnarray}

\subsection{Retarded and advanced Green's functions}

Using the previously obtained eigensystem of $K$, the retarded and
advanced components of the Green's function for $x,x'\in I_{0}$ are
given by 

\begin{eqnarray}
\bra x\bs G^{R}\left(\omega\right)\ket{x'} & = & \frac{1}{2L}\sum_{n}\left(\begin{array}{c}
e^{i\phi_{\text{L}}}e^{-2a\gamma_{\text{L}}}e^{iq_{n}x}\\
e^{-iq_{n}x}
\end{array}\right)\frac{1}{\omega-vq_{n}}\left(\begin{array}{cc}
e^{-i\phi_{\text{L}}}e^{2a\gamma_{\text{L}}}e^{-iq_{n}x'} & e^{iq_{n}x'}\end{array}\right)
\end{eqnarray}
and $\bs G^{A}\left(\omega\right)=\left[\bs G^{R}\left(\omega\right)\right]^{\dagger}$.
The sum over the quantized $q_{n}$'s can be replaced by the contour
integral 
\begin{eqnarray}
\frac{1}{2L}\sum_{n}\frac{e^{iq_{n}x}}{\omega-vq_{n}} & = & \varointctrclockwise_{z_{i}}\frac{dq}{2\pi}\frac{e^{iq_{n}x}}{\omega-vq_{n}}z_{\pm}\left(q\right)=\frac{i}{v}z_{\pm}\left(\omega v^{-1}\right)e^{i\omega v^{-1}x},
\end{eqnarray}
where
\begin{eqnarray}
z_{\pm}\left(q\right) & = & \frac{\pm1}{e^{\pm\left[2iqL-2a\left(\gamma_{\text{L}}+\gamma_{\text{R}}\right)+i\left(\phi_{\text{L}}-\phi_{\text{R}}\right)\right]}-1},
\end{eqnarray}
for $x>0$ or $x<0$ respectively, can be chosen in order to render
the integral convergent once the contour is deformed. Using this identity
we obtain 
\begin{eqnarray}
\bra x\bs G^{R}\left(\omega\right)\ket{x'} & = & i\frac{1}{v}\left[\Theta\left(x-x'\right)\left(\begin{array}{cc}
z_{+}\left(\omega v^{-1}\right)e^{i\omega v^{-1}\left(x-x'\right)} & e^{i\phi_{\text{L}}}e^{-2a\gamma_{\text{L}}}z_{+}\left(\omega v^{-1}\right)e^{i\omega v^{-1}\left(x+x'\right)}\\
e^{-i\phi_{\text{L}}}e^{2a\gamma_{\text{L}}}z_{-}\left(\omega v^{-1}\right)e^{-i\omega v^{-1}\left(x+x'\right)} & z_{-}\left(\omega v^{-1}\right)e^{-i\omega v^{-1}\left(x-x'\right)}
\end{array}\right)\right.\nonumber \\
 &  & \left.+\Theta\left(x'-x\right)\left(\begin{array}{cc}
z_{-}\left(\omega v^{-1}\right)e^{i\omega v^{-1}\left(x-x'\right)} & e^{i\phi_{\text{L}}}e^{-2a\gamma_{\text{L}}}z_{+}\left(\omega v^{-1}\right)e^{i\omega v^{-1}\left(x+x'\right)}\\
e^{-i\phi_{\text{L}}}e^{2a\gamma_{\text{L}}}z_{-}\left(\omega v^{-1}\right)e^{-i\omega v^{-1}\left(x+x'\right)} & z_{+}\left(\omega v^{-1}\right)e^{-i\omega v^{-1}\left(x-x'\right)}
\end{array}\right)\right]\nonumber \\
\end{eqnarray}
The spectral function is thus given by 
\begin{eqnarray}
\bra x\bs{\rho}^{-}\left(\omega\right)\ket{x'} & = & \frac{1}{2\pi v}\frac{\sinh\left(2a\gamma_{\text{L}}\right)}{\sinh\left[2a\left(\gamma_{\text{L}}+\gamma_{\text{R}}\right)\right]}\left(\begin{array}{cc}
e^{2a\gamma_{\text{R}}}u_{L,0}\left(\omega v^{-1}\right)e^{i\omega v^{-1}\left(x-x'\right)} & e^{i\phi_{\text{L}}}u_{L,-2}\left(\omega v^{-1}\right)e^{i\omega v^{-1}\left(x+x'\right)}\\
e^{-i\phi_{\text{L}}}u_{L,2}\left(\omega v^{-1}\right)e^{-i\omega v^{-1}\left(x+x'\right)} & e^{-2a\gamma_{\text{R}}}u_{L,0}\left(\omega v^{-1}\right)e^{-i\omega v^{-1}\left(x-x'\right)}
\end{array}\right)\nonumber \\
 &  & \frac{1}{2\pi v}\frac{\sinh\left(2a\gamma_{\text{R}}\right)}{\sinh\left[2a\left(\gamma_{\text{L}}+\gamma_{\text{R}}\right)\right]}\left(\begin{array}{cc}
e^{-2a\gamma_{\text{L}}}u_{L,0}\left(\omega v^{-1}\right)e^{i\omega v^{-1}\left(x-x'\right)} & e^{i\phi_{\text{L}}}u_{L,0}\left(\omega v^{-1}\right)e^{i\omega v^{-1}\left(x+x'\right)}\\
e^{-i\phi_{\text{L}}}u_{L,0}\left(\omega v^{-1}\right)e^{-i\omega v^{-1}\left(x+x'\right)} & e^{2a\gamma_{L}}u_{L,0}\left(\omega v^{-1}\right)e^{-i\omega v^{-1}\left(x-x'\right)}
\end{array}\right)\nonumber \\
\label{eq:rho_m}
\end{eqnarray}
where we defined 
\begin{eqnarray}
u_{L,n}\left(q\right) & = & -\frac{e^{i\left[qL+\frac{1}{2}\left(\phi_{\text{L}}-\phi_{\text{R}}\right)\right]n}\sinh\left[2a\left(\gamma_{\text{L}}+\gamma_{\text{R}}\right)\right]}{\cos\left[2qL+\left(\phi_{\text{L}}-\phi_{\text{R}}\right)\right]-\cosh\left[2a\left(\gamma_{\text{L}}+\gamma_{\text{R}}\right)\right]}.
\end{eqnarray}
In the limit $L\to\infty$ one has that, for a regular function $f\left(q\right)$
and $\epsilon>0$ independent of $L$, 
\begin{eqnarray}
\lim_{L\to\infty}\frac{1}{2\epsilon}\int_{q-\epsilon}^{q+\epsilon}dq'\ u_{L,n}\left(q'\right)f\left(q'\right) & = & u_{n}\ f\left(q\right)\label{eq:averaging}
\end{eqnarray}
with 
\begin{eqnarray}
u_{0} & = & 1\\
u_{\pm1} & = & 0\\
u_{\pm2} & = & e^{-2a\left(\gamma_{\text{L}}+\gamma_{\text{R}}\right)}
\end{eqnarray}
Using these limiting identities, for $\abs x,\abs{x'}\ll L/2$ and
$L\to\infty$, we can obtain the behavior of the spectral function,
deep into the wire or closer to the boundaries: 
\begin{eqnarray}
\bra x\bs{\rho}_{\text{bulk}}^{-}\left(\omega\right)\ket{x'} & \equiv & \bra{x+L/2}\bs{\rho}^{-}\left(\omega\right)\ket{x'+L/2}=\frac{1}{2\pi v}\left(\begin{array}{cc}
e^{i\omega v^{-1}\left(x-x'\right)} & 0\\
0 & e^{-i\omega v^{-1}\left(x-x'\right)}
\end{array}\right)\label{eq:rho_m_b}\\
\bra x\bs{\rho}_{\text{left}}^{-}\left(\omega\right)\ket{x'} & \equiv & \bra x\bs{\rho}^{-}\left(\omega\right)\ket{x'}=\frac{1}{2\pi v}\left(\begin{array}{cc}
e^{i\omega v^{-1}\left(x-x'\right)} & e^{i\phi_{\text{L}}-2a\gamma_{\text{L}}}e^{i\omega v^{-1}\left(x+x'\right)}\\
e^{-i\phi_{\text{L}}-2a\gamma_{\text{L}}}e^{-i\omega v^{-1}\left(x+x'\right)} & e^{-i\omega v^{-1}\left(x-x'\right)}
\end{array}\right)\label{eq:rho_m_l}\\
\bra x\bs{\rho}_{\text{right}}^{-}\left(\omega\right)\ket{x'} & \equiv & \bra{x+L}\bs{\rho}^{-}\left(\omega\right)\ket{x'+L}=\frac{1}{2\pi v}\left(\begin{array}{cc}
e^{i\omega v^{-1}\left(x-x'\right)} & e^{i\phi_{\text{R}}-2a\gamma_{\text{R}}}e^{i\omega v^{-1}\left(x+x'\right)}\\
e^{-i\phi_{\text{R}}-2a\gamma_{\text{R}}}e^{-i\omega v^{-1}\left(x+x'\right)} & e^{-i\omega v^{-1}\left(x-x'\right)}
\end{array}\right)\label{eq:rho_m_r}
\end{eqnarray}

\subsection{Keldysh Green's functions}

In the steady state, the Keldysh component of the Green's function
is given by $G^{K}=G^{R}\Sigma^{K}G^{A}$ with $\Sigma^{K}=\Sigma_{\text{L}}^{K}+\Sigma_{\text{R}}^{K}$
and 

\begin{eqnarray}
\bra x\bs{\Sigma}_{l}^{K}\left(\omega\right)\ket{x'} & = & -2iv\,\gamma_{l}\Theta\left(\abs{x-x_{l}}-a\right)\delta\left(x-x'\right)\tanh\left[\frac{\beta_{l}}{2}\left(\omega-\mu_{l}\right)\right]\sigma_{0},
\end{eqnarray}
this yields 
\begin{eqnarray}
\bra x\bs G^{K}\left(\omega\right)\ket{x'} & = & \sum_{l=\text{L},\text{R}}\int_{I_{l}}dy\,\bra x\bs G^{R}\left(\omega\right)\ket y\bra y\bs{\Sigma}_{l}^{K}\left(\omega\right)\ket y\bra y\bs G^{A}\left(\omega\right)\ket{x'}.
\end{eqnarray}
For $x,x'\in I_{0}$ one gets 

\begin{eqnarray}
 &  & \bra x\bs{\rho}^{+}\left(\omega\right)\ket{x'}=\label{eq:rho_p}\\
 &  & \frac{1}{2\pi v}\tanh\left[\frac{\beta_{\text{L}}}{2}\left(\omega-\mu_{\text{L}}\right)\right]\frac{\sinh\left(2a\gamma_{\text{L}}\right)}{\sinh\left[2a\left(\gamma_{\text{L}}+\gamma_{\text{R}}\right)\right]}\left(\begin{array}{cc}
e^{2a\gamma_{\text{R}}}u_{L,0}\left(\omega v^{-1}\right)e^{i\omega v^{-1}\left(x-x'\right)} & u_{L,-2}\left(\omega v^{-1}\right)e^{i\phi_{\text{L}}}e^{i\omega v^{-1}\left(x+x'\right)}\\
u_{L,2}\left(\omega v^{-1}\right)e^{-i\phi_{\text{L}}}e^{-i\omega v^{-1}\left(x+x'\right)} & e^{-2a\gamma_{\text{R}}}u_{L,0}\left(\omega v^{-1}\right)e^{-i\omega v^{-1}\left(x-x'\right)}
\end{array}\right)\nonumber \\
 &  & \frac{1}{2\pi v}\tanh\left[\frac{\beta_{\text{R}}}{2}\left(\omega-\mu_{\text{R}}\right)\right]\frac{\sinh\left(2a\gamma_{\text{R}}\right)}{\sinh\left[2a\left(\gamma_{\text{L}}+\gamma_{\text{R}}\right)\right]}\left(\begin{array}{cc}
u_{L,0}\left(\omega v^{-1}\right)e^{-2a\gamma_{\text{L}}}e^{i\omega v^{-1}\left(x-x'\right)} & u_{L,0}\left(\omega v^{-1}\right)e^{i\phi_{\text{L}}}e^{i\omega v^{-1}\left(x+x'\right)}\\
e^{-i\phi_{\text{L}}}u_{L,0}\left(\omega v^{-1}\right)e^{-i\omega v^{-1}\left(x+x'\right)} & e^{2a\gamma_{\text{L}}}u_{L,0}\left(\omega v^{-1}\right)e^{-i\omega v^{-1}\left(x-x'\right)}
\end{array}\right).\nonumber 
\end{eqnarray}
For $\abs x,\abs{x'}\ll L/2$ and $L\to\infty$, we obtain 
\begin{eqnarray}
 &  & \bra x\bs{\rho}_{\text{bulk}}^{+}\left(\omega\right)\ket{x'}\equiv\bra{x+L/2}\bs{\rho}^{+}\left(\omega\right)\ket{x'+L/2}=\nonumber \\
 &  & \frac{1}{2\pi v}\tanh\left[\frac{\beta_{\text{L}}}{2}\left(\omega-\mu_{\text{L}}\right)\right]\frac{\sinh\left(2a\gamma_{\text{L}}\right)}{\sinh\left[2a\left(\gamma_{\text{L}}+\gamma_{\text{R}}\right)\right]}\left(\begin{array}{cc}
e^{2a\gamma_{\text{R}}}e^{i\omega v^{-1}\left(x-x'\right)} & 0\\
0 & e^{-2a\gamma_{\text{R}}}e^{-i\omega v^{-1}\left(x-x'\right)}
\end{array}\right)\nonumber \\
 &  & +\frac{1}{2\pi v}\tanh\left[\frac{\beta_{\text{R}}}{2}\left(\omega-\mu_{\text{R}}\right)\right]\frac{\sinh\left(2a\gamma_{\text{R}}\right)}{\sinh\left[2a\left(\gamma_{\text{L}}+\gamma_{\text{R}}\right)\right]}\left(\begin{array}{cc}
e^{-2a\gamma_{\text{L}}}e^{i\omega v^{-1}\left(x-x'\right)} & 0\\
0 & e^{2a\gamma_{\text{L}}}e^{-i\omega v^{-1}\left(x-x'\right)}
\end{array}\right);\nonumber \\
\label{eq:rho_p_b}\\
 &  & \bra x\bs{\rho}_{\text{left}}^{+}\left(\omega\right)\ket{x'}\equiv\bra x\bs{\rho}^{+}\left(\omega\right)\ket{x'}=\nonumber \\
 &  & \frac{1}{2\pi v}\tanh\left[\frac{\beta_{L}}{2}\left(\omega-\mu_{\text{L}}\right)\right]\frac{\sinh\left(2a\gamma_{\text{L}}\right)}{\sinh\left[2a\left(\gamma_{\text{L}}+\gamma_{\text{R}}\right)\right]}\left(\begin{array}{cc}
e^{2a\gamma_{\text{R}}}e^{i\omega v^{-1}\left(x-x'\right)} & e^{-2\left(a\gamma_{\text{L}}+a\gamma_{\text{R}}\right)}e^{i\phi_{\text{L}}}e^{i\omega v^{-1}\left(x+x'\right)}\\
e^{-2\left(a\gamma_{\text{L}}+a\gamma_{\text{R}}\right)}e^{-i\phi_{\text{L}}}e^{-i\omega v^{-1}\left(x+x'\right)} & e^{-2a\gamma_{\text{R}}}e^{-i\omega v^{-1}\left(x-x'\right)}
\end{array}\right)\nonumber \\
 &  & +\frac{1}{2\pi v}\tanh\left[\frac{\beta_{\text{R}}}{2}\left(\omega-\mu_{\text{R}}\right)\right]\frac{\sinh\left(2a\gamma_{\text{R}}\right)}{\sinh\left[2a\left(\gamma_{\text{L}}+\gamma_{\text{R}}\right)\right]}\left(\begin{array}{cc}
e^{-2a\gamma_{\text{L}}}e^{i\omega v^{-1}\left(x-x'\right)} & e^{i\phi_{\text{L}}}e^{i\omega v^{-1}\left(x+x'\right)}\\
e^{-i\phi_{\text{L}}}e^{-i\omega v^{-1}\left(x+x'\right)} & e^{2a\gamma_{\text{L}}}e^{-i\omega v^{-1}\left(x-x'\right)}
\end{array}\right);\nonumber \\
\label{eq:rho_p_l}\\
 &  & \bra x\bs{\rho}_{\text{right}}^{+}\left(\omega\right)\ket{x'}\equiv\bra{x+L}\bs{\rho}^{+}\left(\omega\right)\ket{x'+L}=\nonumber \\
 &  & \frac{1}{2\pi v}\tanh\left[\frac{\beta_{\text{L}}}{2}\left(\omega-\mu_{\text{L}}\right)\right]\frac{\sinh\left(2a\gamma_{\text{L}}\right)}{\sinh\left[2a\left(\gamma_{\text{L}}+\gamma_{\text{R}}\right)\right]}\left(\begin{array}{cc}
e^{2a\gamma_{\text{R}}}e^{i\omega v^{-1}\left(x-x'\right)} & e^{i\phi_{\text{R}}}e^{i\omega v^{-1}\left(x+x'\right)}\\
e^{-i\phi_{\text{R}}}e^{-i\omega v^{-1}\left(x+x'\right)} & e^{-2a\gamma_{\text{R}}}e^{-i\omega v^{-1}\left(x-x'\right)}
\end{array}\right)\nonumber \\
 &  & +\frac{1}{2\pi v}\tanh\left[\frac{\beta_{\text{R}}}{2}\left(\omega-\mu_{\text{R}}\right)\right]\frac{\sinh\left(2a\gamma_{\text{R}}\right)}{\sinh\left[2a\left(\gamma_{\text{L}}+\gamma_{\text{R}}\right)\right]}\left(\begin{array}{cc}
e^{-2a\gamma_{\text{L}}}e^{i\omega v^{-1}\left(x-x'\right)} & e^{-2\left(a\gamma_{\text{L}}+a\gamma_{\text{R}}\right)}e^{i\phi_{\text{R}}}e^{i\omega v^{-1}\left(x+x'\right)}\\
e^{-2\left(a\gamma_{\text{L}}+a\gamma_{\text{R}}\right)}e^{-i\phi_{\text{R}}}e^{-i\omega v^{-1}\left(x+x'\right)} & e^{2a\gamma_{\text{L}}}e^{-i\omega v^{-1}\left(x-x'\right)}
\end{array}\right);\nonumber \\
\label{eq:rho_p_r}
\end{eqnarray}

\subsection{Bulk in the $L\to\infty$ limit}

In the limit $L\to\infty$, the correlation functions within the bulk
region becomes translational invariant $\bra x\bs{\rho}_{\text{bulk}}^{\pm}\left(\omega\right)\ket{x'}=\bra{x-y}\bs{\rho}_{\text{bulk}}^{\pm}\left(\omega\right)\ket{x'-y}$.
One can thus consider the Fourier transformed quantities: 
\begin{eqnarray}
\bs{\rho}_{\text{bulk }}^{\pm}\left(\omega,q\right) & = & \int dxdx'\,e^{-iq\left(x-x'\right)}\bra x\bs{\rho}_{\text{bulk}}^{\pm}\left(\omega\right)\ket{x'}.
\end{eqnarray}
Explicitly we obtain 
\begin{eqnarray}
\bs{\rho}_{\text{bulk }}^{\pm}\left(\omega,q\right) & = & \left(\begin{array}{cc}
\rho_{\text{L }}^{\pm}\left(\omega,q\right) & 0\\
0 & \rho_{\text{R }}^{\pm}\left(\omega,q\right)
\end{array}\right),
\end{eqnarray}
with 
\begin{eqnarray*}
\rho_{\text{L }}^{-}\left(\omega,q\right) & = & \delta\left(\omega-vq\right),\\
\rho_{\text{R }}^{-}\left(\omega,q\right) & = & \delta\left(\omega+vq\right),\\
\rho_{\text{L }}^{+}\left(\omega,q\right) & = & F_{\text{L}}\left(\omega\right)\delta\left(\omega-vq\right),\\
\rho_{\text{R }}^{+}\left(\omega,q\right) & = & F_{\text{R}}\left(\omega\right)\delta\left(\omega+vq\right),
\end{eqnarray*}
where 
\begin{eqnarray}
F_{\text{L}}\left(\omega\right) & = & b_{\text{L}}\tanh\left[\frac{\beta_{\text{L}}}{2}\left(\omega-\mu_{\text{L}}\right)\right]+\left(1-b_{\text{L}}\right)\tanh\left[\frac{\beta_{\text{R}}}{2}\left(\omega-\mu_{\text{R}}\right)\right]\\
F_{\text{R}}\left(\omega\right) & = & \left(1-b_{\text{R}}\right)\tanh\left[\frac{\beta_{\text{L}}}{2}\left(\omega-\mu_{\text{L}}\right)\right]+b_{\text{R}}\tanh\left[\frac{\beta_{\text{R}}}{2}\left(\omega-\mu_{\text{R}}\right)\right]
\end{eqnarray}
and 
\begin{eqnarray*}
e^{4a\gamma_{\text{L}}} & = & \frac{1-\mathit{b}_{\text{R}}}{1-\mathit{b}_{\text{L}}},\\
e^{4a\gamma_{\text{R}}} & = & \frac{\mathit{b}_{\text{L}}}{\mathit{b}_{\text{R}}}.
\end{eqnarray*}

For the tight binding model, with $r$ and $r'$ in the middle of
the wire, we obtain
\begin{eqnarray}
\rho_{rr'}^{\pm}\left(\omega\right) & \simeq & a_{0}\int_{-\Lambda}^{\Lambda}\frac{dq}{2\pi}\left[\rho_{\text{L }}^{\pm}\left(\omega,q\right)e^{iqa_{0}\left(r-r'\right)}e^{ik_{\text{F}}\left(r-r'\right)}+\rho_{\text{R }}^{\pm}\left(\omega,q\right)e^{iqa_{0}\left(r-r'\right)}e^{-ik_{\text{F}}\left(r-r'\right)}\right],
\end{eqnarray}
or inverting the Fourier transform
\begin{eqnarray}
\rho_{k}^{\pm}\left(\omega\right) & = & \rho_{\text{L }}^{\pm}\left[\omega,\left(k-k_{\text{F}}\right)a_{0}^{-1}\right]\Theta\left(\abs{k-k_{\text{F}}}<\Lambda a_{0}\right)+\rho_{\text{R }}^{\pm}\left(\omega,k+k_{\text{F}}\right)\Theta\left(\abs{k+k_{\text{F}}}<\Lambda a_{0}\right).
\end{eqnarray}
where $k\in\left[0,2\pi\right]$ . 

The single particle density matrix, defined by 
\begin{eqnarray}
\varrho_{rr'}\left(t\right) & = & \av{c_{r}^{\dagger}\left(t\right)c_{r'}\left(t\right)},
\end{eqnarray}
that in the steady state is given by 
\begin{eqnarray*}
\varrho_{rr'} & = & -\pi\int\frac{d\omega}{2\pi}\rho_{r'r}^{+}\left(\omega\right)+\frac{1}{2}\delta_{rr'},
\end{eqnarray*}
can be approximated by 
\begin{eqnarray}
\varrho_{rr'} & = & \int_{-\pi}^{\pi}\frac{dk}{2\pi}e^{-i\left(r-r'\right)k}n_{k}
\end{eqnarray}
with $n_{k}$ the occupation number of the mode $k$ given by
\begin{eqnarray*}
n_{k} & \simeq & \begin{cases}
1 & -k_{F}+a_{0}\Lambda<k<k_{F}-a_{0}\Lambda;\\
\frac{1}{2}\left\{ 1-F_{\text{L}}\left[\frac{v}{a_{0}}\left(k-k_{F}\right)\right]\right\}  & \text{for }k_{F}-a_{0}\Lambda<k<k_{F}+a_{0}\Lambda;\\
0 & k_{F}+a_{0}\Lambda<k\vee k<-k_{F}-a_{0}\Lambda;\\
\frac{1}{2}\left\{ 1-F_{\text{R}}\left[\frac{v}{a_{0}}\left(-k_{F}-k\right)\right]\right\}  & \text{for }-k_{F}-a_{0}\Lambda<k<-k_{F}+a_{0}\Lambda;
\end{cases}.
\end{eqnarray*}

\section{Identifications with the Tight-Binding model II - $\phi_{l}$ and
$\gamma_{l}$ }

\subsection{Self-energies}

In the tight-binding model the self-energy contribution due to the
reservoirs is given by 

\begin{eqnarray*}
\Sigma_{\text{TB},rr'}^{R/A}\left(\omega\right) & = & \mp i\sum_{l}\Sigma_{\text{TB},rr';l}^{R/A}\left(\omega\right)
\end{eqnarray*}
where 
\begin{eqnarray*}
\Sigma_{\text{TB},rr';l}^{R/A}\left(\omega\right) & = & \mp i\left(\Gamma_{l}\delta_{r'r_{l}}\delta_{rr_{l}}\right)
\end{eqnarray*}
is the contribution from the reservoir $l$ and $\Gamma_{l}$ the
hybridization constant \cite{Ribeiro2015a}. The Keldysh component
writes 
\begin{eqnarray*}
\Sigma_{\text{TB},rr'}^{K}\left(\omega\right) & = & \sum_{l}\tanh\left[\frac{\beta_{l}}{2}\left(\omega-\mu_{l}\right)\right]\left[\Sigma_{\text{TB},rr',l}^{R}\left(\omega\right)-\Sigma_{\text{TB},rr',l}^{A}\left(\omega\right)\right].
\end{eqnarray*}

\subsection{Determination of $\phi_{l}$ }

The tight-binding Hamiltonian, given by 
\begin{align}
\bs H_{\text{TB}} & =-t\left[\left(\sum_{r=0}^{N-2}\ket r\bra{r+1}\right)+\text{h.c.}\right],
\end{align}
has wave functions of the from 
\begin{eqnarray}
\psi\left(r\right)=\braket r{\psi_{k}} & = & Ae^{ikr}+Be^{-ikr},
\end{eqnarray}
and the dispersion relation 
\begin{eqnarray}
\varepsilon_{k} & = & -2\cos\left(k\right).
\end{eqnarray}
Quantization condition for the momentum can be obtained imposing $\bra{r=0}\bs H_{\text{TB}}\ket{\psi_{k}}=\varepsilon_{k}\braket 0{\psi_{k}}$
and $\bra{r=N-1}\bs H_{\text{TB}}\ket{\psi_{k}}=\varepsilon_{k}\braket{N-1}{\psi_{k}}$
and are equivalent to the relations

\begin{align}
\psi\left(-1\right) & =\psi\left(N\right)=0.
\end{align}
For the wave function expanded around $k_{\text{F}}$ 
\begin{equation}
\braket r{\psi_{k}}\simeq e^{ik_{F}r}\braket{x=a_{0}r}{\psi_{\text{L}}}+e^{-ik_{F}r}\braket{x=a_{0}r}{\psi_{\text{R}}},
\end{equation}
this implies, in the limit $a_{0}\to0$, $a_{0}N\to L$, 
\begin{eqnarray}
e^{-ik_{\text{F}}}\braket 0{\psi_{\text{L}}}+e^{ik_{\text{F}}r}\braket 0{\psi_{\text{R}}} & = & 0,\\
e^{ik_{\text{F}}La_{0}^{-1}}\braket L{\psi_{\text{L}}}+e^{-ik_{\text{F}}La_{0}^{-1}}\braket L{\psi_{\text{R}}} & = & 0,
\end{eqnarray}
yielding to the phase shifts 
\begin{align}
\phi_{L} & =2k_{\text{F}}-\pi,\\
\phi_{R} & =-2k_{\text{F}}La_{0}^{-1}-\pi.
\end{align}

\subsection{Determination of $\gamma_{l}$ }

The eigensystem of
\begin{align*}
\bs K_{\text{TB}} & =-t\left[\left(\sum_{r=1}^{N-1}\ket r\bra{r+1}\right)+\text{h.c.}\right]-i\left(\Gamma_{\text{L}}\ket 1\bra 1+\Gamma_{\text{R}}\ket N\bra N\right),
\end{align*}
that is the tight binding version of the $\bs K$ operator, can be
obtained in a similar way to the one of the previous section. 

The spectrum of $\bs K_{\text{TB}}$ is obtained imposing $\bra{r=0}\bs H_{\text{TB}}\ket{\psi_{k}}=\varepsilon_{k}\braket 0{\psi_{k}}$
and $\bra{r=N-1}\bs H_{\text{TB}}\ket{\psi_{k}}=\varepsilon_{k}\braket{N-1}{\psi_{k}}$,
and yields to 
\begin{eqnarray*}
\left[\frac{te^{-ik}-i\Gamma_{\text{L}}}{te^{ik}-i\Gamma_{\t L}}\right]\left[\frac{te^{-ik}-i\Gamma_{\t R}}{te^{ik}-i\Gamma_{\t R}}\right] & = & e^{2ikN}.
\end{eqnarray*}
Assuming $k=k_{0}+\frac{\Delta k}{N}$ where $k_{0}$ is a solution
of the equation 
\begin{align}
e^{-4ik_{0}} & =e^{2ik_{0}N},
\end{align}
i..e. 
\begin{align}
k_{0} & =\frac{\pi}{L+1}n
\end{align}
and $\Delta k$ is of order $1/N$, we obtain 
\begin{eqnarray}
\Delta k & = & \frac{1}{2}\tan^{-1}\left\{ \frac{\sin\left(2k_{0}\right)\left[\left(\Gamma_{\t R}^{2}+\Gamma_{\t L}^{2}\right)+2\cos\left(2k_{0}\right)\Gamma_{\t L}^{2}\Gamma_{\t R}^{2}\right]}{\Gamma_{\t L}^{2}\Gamma_{\t R}^{2}\cos\left(4k_{0}\right)+\left(\Gamma_{\t L}^{2}+\Gamma_{\t R}^{2}\right)\cos\left(2k_{0}\right)+t^{2}}\right\} \\
 &  & +\frac{1}{4}i\left[\ln\left(-\frac{\Gamma_{\t L}^{2}t^{-2}-2t^{-1}\sin k_{0}\Gamma_{\t L}+1}{\Gamma_{\t L}^{2}t^{-2}+2t^{-1}\sin k_{0}\Gamma_{\t L}+1}\right)+\ln\left(-\frac{\Gamma_{\t R}^{2}t^{-2}-2t^{-1}\sin k_{0}\Gamma_{\t R}+1}{\Gamma_{\t R}^{2}t^{-2}+2t^{-1}\sin k_{0}\Gamma_{\t R}+1}\right)\right].\nonumber 
\end{eqnarray}
Therefore near $k_{0}=k_{\t F}$ we get
\begin{align}
\im q & =a_{0}^{-1}\im\left(k_{0}-k_{F}\right)\\
 & =-\frac{1}{4}i\frac{1}{a_{0}N}\left[\ln\left(-\frac{\Gamma_{\t L}^{2}t^{-2}+2t^{-1}\sin k_{\t F}\Gamma_{\t L}+1}{\Gamma_{\t L}^{2}t^{-2}-2t^{-1}\sin k_{\t F}\Gamma_{\t L}+1}\right)+\ln\left(-\frac{\Gamma_{\t R}^{2}t^{-2}+2t^{-1}\sin k_{\t F}\Gamma_{\t R}+1}{\Gamma_{\t R}^{2}t^{-2}-2t^{-1}\sin k_{\t F}\Gamma_{\t R}+1}\right)\right].
\end{align}
Identifying this expression with the quantization condition of the
continuum model given by Eq.(\ref{eq:quantization_cond}) we obtain
\begin{align}
a\gamma_{l} & =\frac{1}{4}\ln\left(-\frac{\Gamma_{l}^{2}t^{-2}-2t^{-1}\sin k_{\t F}\Gamma_{l}+1}{\Gamma_{l}^{2}t^{-2}+2t^{-1}\sin k_{\t F}\Gamma_{l}+1}\right)
\end{align}
 that for $\Gamma_{l}\ll1$ can be approximated by $a\gamma_{l}\simeq\sin k_{\t F}\Gamma_{l}$. 

\subsection{Numerical results for the two-point function }

Fig. \ref{fig:One-point-functions} shows the one-point functions
$\rho_{rr'}^{\pm}\left(\omega\right)$ as a function of $\omega$
for a finite system. The sharp features with frequency are signatures
of the discrete spectrum of the finite chain in the absence of the
leads and, in the spectral function $\rho_{rr'}^{-}\left(\omega\right)$,
they become $\delta$-functions in the limit $\Gamma_{\t L},\Gamma_{\t R}\to0$.
The infinite volume approximation washes out the rapid variations
by averaging over small energy window of the order of the level spacing
before taking the $L\to\infty$ limit, see Eq.(\ref{eq:averaging}).

\begin{figure}[H]
\centering{}\includegraphics[width=0.7\textwidth]{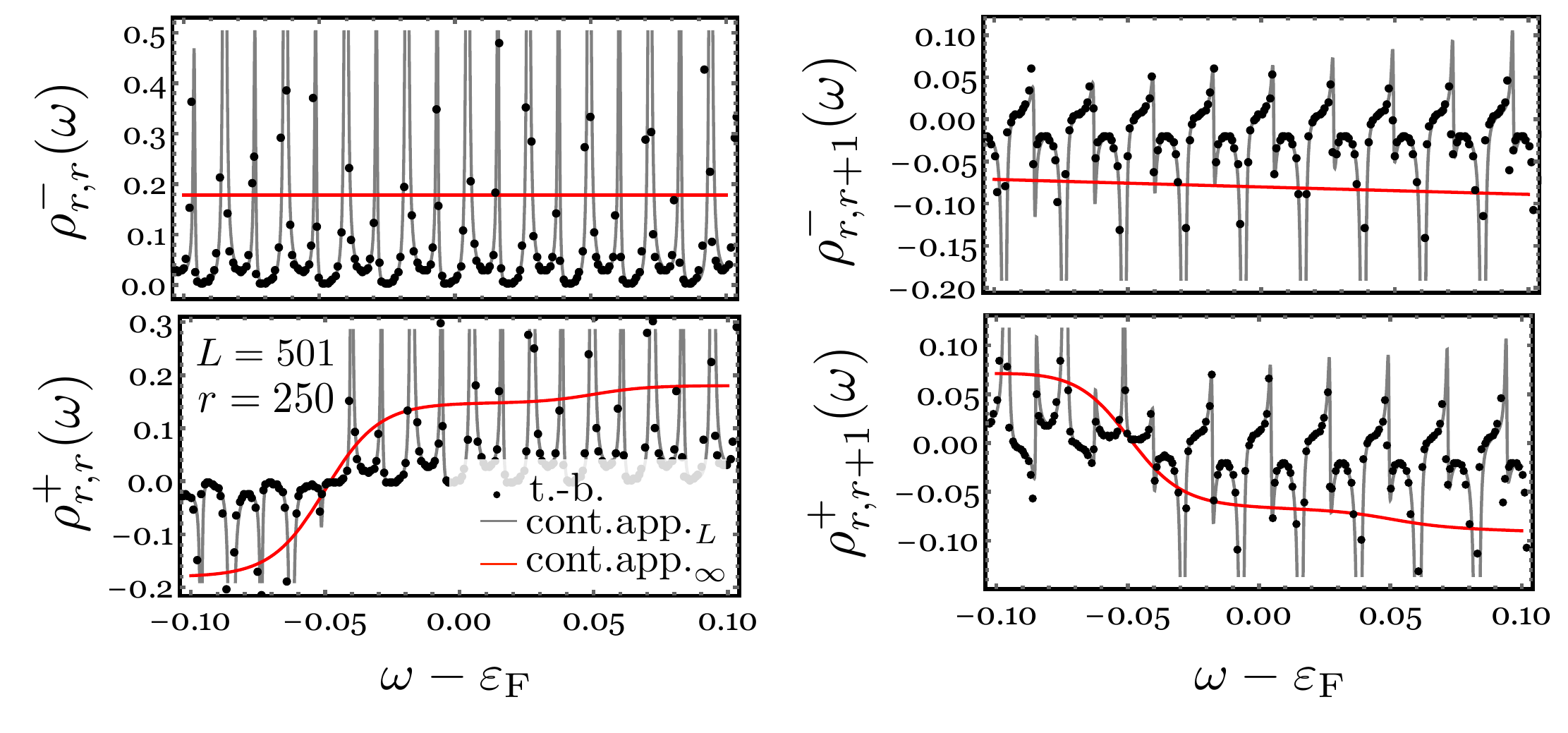}\caption{\label{fig:One-point-functions} One-point functions $\rho_{rr'}^{\pm}\left(\omega\right)$
as a function of $\omega$ computed for $r'=r$ (left panel) and $r'=r+1$
(right panel) and for $r=250$, $L=501$, $\Gamma_{\protect\t L}=0.5t$,
$\Gamma_{\protect\t R}=0.2t$, $\varepsilon_{\protect\t F}=0.9t$,
$V=0.1t$, and $T_{\protect\t L}=T_{\protect\t R}=0.01t$. The black
dots are obtained numerically using the tight-binding model. The black
curves are obtained from the continuum limit for finite $L$ using
Eqs. (\ref{eq:rho_m},\ref{eq:rho_p}). The infinite volume limit
is obtained form Eqs.(\ref{eq:rho_m_b}-\ref{eq:rho_m_r}) and Eqs.(\ref{eq:rho_p_b}-\ref{eq:rho_p_r}). }
\end{figure}

\section{Entanglement entropy}

\subsection{Derivation}

The entanglement entropy of a region $\Sigma$ is defined as 

\begin{eqnarray}
S_{\Sigma} & = & -\tr\left(\hat{\rho}_{\Sigma}\ln\hat{\rho}_{\Sigma}\right)
\end{eqnarray}
where $\hat{\rho}_{\Sigma}=\tr_{\bar{\Sigma}}\hat{\rho}$ is obtained
from the total density matrix $\hat{\rho}$ by trancing out the degrees
of freedom belonging $\bar{\Sigma}$, the complement of $\Sigma$.
In the following we consider a region of size $\ell$ in the central
region of the wire 
\begin{eqnarray}
\Sigma_{\ell} & = & \left\{ r:\ r\in\left[\left(L-\ell\right)/2,\left(L+\ell\right)/2\right]\right\} ,
\end{eqnarray}
and we denote $S_{\ell}=S_{\Sigma_{\ell}}$. Since the model is Gaussian
we have 
\begin{eqnarray}
S_{\ell} & = & \tr\left[s\left(\varrho_{\ell}\right)\right]
\end{eqnarray}
for 
\begin{align}
s\left(\nu\right) & =-\nu\ln\nu-\left(1-\nu\right)\ln\left(1-\nu\right)
\end{align}
with 
\begin{eqnarray}
\varrho_{\ell} & = & \sum_{rr'\in\Sigma_{\ell}}\ket r\varrho_{rr'}\bra{r'}
\end{eqnarray}
We follow Ref. \cite{Its2009} and write
\begin{eqnarray}
S_{\ell} & = & \frac{1}{2\pi i}\oint dz\,s\left(\nu\right)\tr\left[\frac{1}{z-\varrho_{\ell}}\right]\nonumber \\
 & = & \frac{1}{2\pi i}\oint dz\,s\left(\nu\right)\,\pd_{z}\ln\det\left[z-\varrho_{\ell}\right]
\end{eqnarray}
The determinant can be calculated, asymptotically for $\ell\to\infty$,
following Ref. \cite{Its2009} and using known results for approximating
the determinant $D_{\ell}\left[\phi\right]=\det\left[z-\varrho_{\ell}\right]$
of the Toplitz-like matrix $z-\varrho_{\ell}$ with symbol 
\begin{eqnarray}
\phi\left(k\right) & = & z-n_{k}.
\end{eqnarray}
For the case $T_{R},T_{L}\neq0$ the symbol $\phi$ is smooth everywhere
and thus the Szegö limit theorem yields to
\begin{eqnarray}
D_{\ell}\left[\phi\right] & \simeq & E_{\text{Sz}}\left[\phi\right]e^{\ell\int\frac{d\theta}{2\pi}\ln\phi\left(\theta\right)},
\end{eqnarray}
where $E_{\text{Sz}}\left[\phi\right]$ is an $\ell$ dependent constant,
which gives 
\begin{eqnarray}
S_{\ell} & = & -\ell\sum_{l}\int_{-\infty}^{\infty}\frac{dk}{2\pi}\,s\left[n_{k}\right]+c^{\text{te}}.
\end{eqnarray}
For the case $T_{R},T_{L}=0$ we have
\begin{eqnarray}
n_{k} & \simeq & \begin{cases}
1 & -k_{F}+a_{0}\Lambda<k<k_{F}-a_{0}\Lambda\\
b_{\text{L}} & \text{for }k_{F}-a_{0}\Lambda<k<k_{F}+a_{0}\Lambda\\
0 & k_{F}+a_{0}\Lambda<k\vee k<-k_{F}-a_{0}\Lambda\\
b_{\text{R}} & \text{for }-k_{F}-a_{0}\Lambda<k<-k_{F}+a_{0}\Lambda
\end{cases}.
\end{eqnarray}
As the symbol $\phi$ is not smooth and we have to use the Fisher-Hartwing
conjecture\cite{Its2009} : 
\begin{eqnarray}
D_{\ell}\left[\phi\right] & \simeq & E_{\text{FH}}\left[\phi\right]e^{\ell V_{0}+\ln\left(\ell\right)\sum_{j=0}^{m}\left(\alpha_{j}^{2}-\beta_{j}^{2}\right)},
\end{eqnarray}
with $E_{\text{FH}}\left[\phi\right]$ an $\ell$ independent constant
where $\beta_{i}$ and $V\left(\theta\right)$ are defined by
\begin{eqnarray}
\phi\left(\theta\right) & = & e^{V\left(\theta\right)}e^{i\sum_{i=0}^{m}\beta_{i}\left(\theta-\theta_{i}\right)}e^{-\pi\sum_{i=0}^{m}\beta_{i}\sgn\left(\theta-\theta_{i}\right)}\abs{2\sin\left(\frac{\theta-\theta_{i}}{2}\right)}^{2\alpha_{j}},
\end{eqnarray}
and 
\begin{eqnarray*}
V_{0} & = & \int\frac{d\theta}{2\pi}V\left(\theta\right),
\end{eqnarray*}
is the regular part. For our case we use the definitions 
\begin{equation}
\theta_{0}=0;\,\theta_{1}=k_{F}-\frac{\Delta}{2};\,\theta_{2}=k_{F}+\frac{\Delta}{2};\,\theta_{3}=2\pi-k_{F}-\frac{\Delta}{2};\,\theta_{4}=2\pi-k_{F}+\frac{\Delta}{2};
\end{equation}
and 
\begin{align}
\beta_{0}=0;\\
\beta_{1}=\frac{1}{2\pi i}\left[\log(z-1)-\log\left(z-\mathit{b}_{R}\right)\right]; & \beta_{2}=\frac{i}{2\pi}\left[\log\left(z-\mathit{b}_{R}\right)-\log(z)\right];\nonumber \\
\beta_{3}=\frac{1}{2\pi i}\left[\log(z)-\log\left(z-\mathit{b}_{L}\right)\right]; & \beta_{4}=\frac{1}{2\pi i}\left[\log\left(z-\mathit{b}_{L}\right)-\log(z-1)\right];\nonumber 
\end{align}
and 
\begin{eqnarray}
e^{V\left(z\right)} & = & (z-1)^{\frac{2k_{F}-\Delta}{2\pi}}z^{\frac{-\Delta-2k_{F}+2\pi}{2\pi}}\left(z-\mathit{b}_{L}\right){}^{\frac{\Delta}{2\pi}}\left(z-\mathit{b}_{R}\right){}^{\frac{\Delta}{2\pi}}.
\end{eqnarray}
The entanglement entropy is therefore given by 
\begin{eqnarray}
S_{\ell} & = & \ell\left[\frac{1}{2\pi i}\oint dz\,s\left(z\right)\pd_{z}V_{0}\right]+\ln\left(\ell\right)\left[\frac{1}{2\pi i}\oint dz\,s\left(z\right)\,\pd_{z}\sum_{j=0}^{m}\left(\alpha_{j}^{2}-\beta_{j}^{2}\right)\right]\label{eq:contour_int}\\
 & = & \frac{\ell\Delta}{2\pi}\left[s\left(b_{L}\right)+s\left(b_{R}\right)\right]+\ln\left(\ell\right)\left[\frac{1}{3}-\tilde{s}\left(b_{L}\right)-\tilde{s}\left(b_{R}\right)\right]+c^{\text{\ensuremath{\underbar{te}}}},
\end{eqnarray}
where 
\begin{eqnarray}
\tilde{s}\left(b\right) & = & \frac{1}{24}+\frac{1}{4\pi^{2}}\left[\left(2\mathit{b}-1\right)\left[\text{Li}_{2}\left(1-\mathit{b}\right)-\text{Li}_{2}\left(\mathit{b}\right)\right]+\left(1-\mathit{b}\right)\log^{2}\left(1-\mathit{b}\right)+\mathit{b}\log^{2}\left(\mathit{b}\right)+\log\left(\mathit{b}\right)\log\left(1-\mathit{b}\right)\right]
\end{eqnarray}
The contour of integration in Eq. (\ref{eq:contour_int}) contains
the segment $z\in[0,1]$. 

\subsection{Additional numerical results}

Fig.\ref{fig:EE} depicts the $\ell$ and the $\ln\ell$ coefficients
of the entanglement entropy $S_{\ell}$, respectively $\gamma_{V}$
and $\tilde{\gamma}_{V}$, as a function of the voltage. The numerical
results obtained for the tight-binding model are seen to converge
to the analytic curves predicted by the continuum model. We observe
that $\gamma_{V}\propto\abs V$ and that $\tilde{\gamma}_{V}$ is
$V$-independent for $V\neq0$. At $V=0$ we recover the equilibrium
result $\tilde{\gamma}_{V}=1/3$.

\begin{figure}[H]
\centering{}\includegraphics[width=0.7\textwidth]{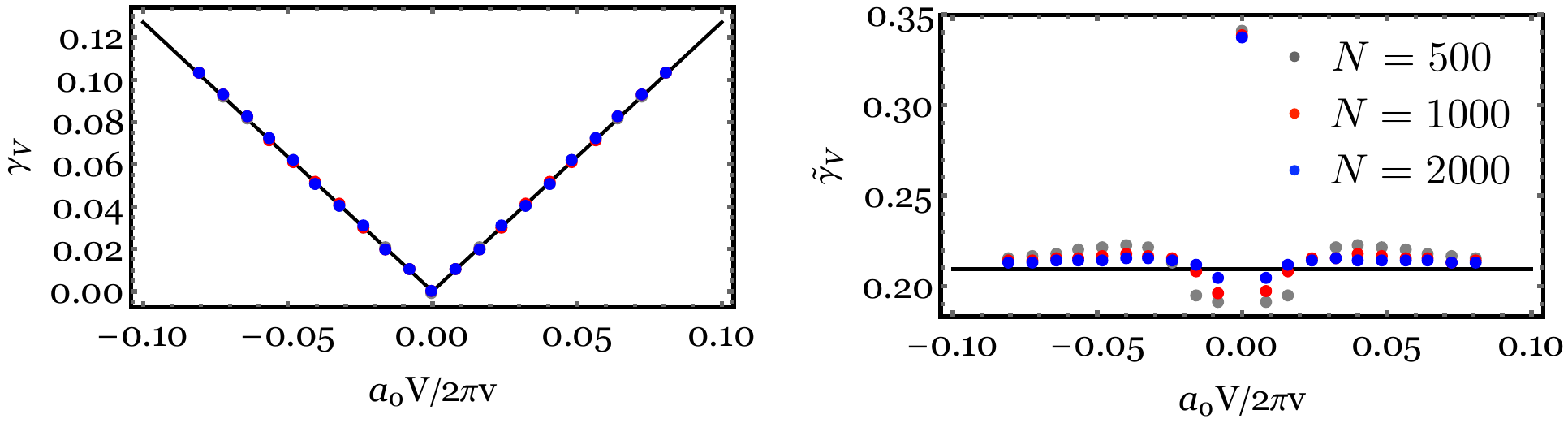}\caption{\label{fig:EE} Volume, $\gamma_{V}$, and logarithmic,$\tilde{\gamma}_{V}$
, coefficients of the entanglement entropy as a function of the applied
voltage $V$ computed for $\Gamma_{\protect\t L}=0.02t$, $\Gamma_{\protect\t R}=0.01t$,
$\varepsilon_{\protect\t F}=0.3t$, and $T_{\protect\t L}=T_{\protect\t R}=0$. }
\end{figure}

\section{Two particle correlation functions}

The charge susceptibility, on the Keldysh contour $\gamma$, is defined
as 
\begin{eqnarray}
\chi\left(zr,z'r'\right) & = & -i\left[\av{T_{\gamma}c_{r}^{\dagger}(z^{+})c_{r}(z)c_{r'}^{\dagger}(z'^{+})c_{r'}(z')}-\av{T_{\gamma}c_{r}^{\dagger}(z^{+})c_{r}(z)}\av{T_{\gamma}c_{r'}^{\dagger}(z'^{+})c_{r'}(z')}\right],
\end{eqnarray}
where the $z^{+}$ denotes a point coming infinitesimally later then
$z$ along $\gamma$. Due to the Gaussian nature of the model
\begin{eqnarray}
\chi\left(zr,z'r'\right) & = & -iG_{r'r}\left(z',z\right)G_{rr'}\left(z',z\right).
\end{eqnarray}
The grater and lesser components are given by $\chi^{>,<}\left(t\bs r,t'\bs r'\right)=\chi\left(z=t\bs r,z'=t'\bs r'\right)$
with $z$ respectively after or before $z'$. For the $r$ and $r'$
in the middle of the wire , using the approximated translational invariant
Green's functions, we obtain: 
\begin{eqnarray}
\chi_{r,r'}^{><}\left(t,t'\right) & = & -i\left[G_{\text{L}}^{<>}\left(t'x',tx\right)G_{\text{L}}^{><}\left(tx,t'x'\right)+G_{\text{R}}^{<>}\left(t'x',tx\right)G_{\text{R}}^{><}\left(tx,t'x'\right)\right]\\
 &  & -i\left[G_{\text{L}}^{<>}\left(t'x',tx\right)G_{\text{R}}^{><}\left(tx,t'x'\right)e^{-i2k_{F}\left(r-r'\right)}+G_{\text{R}}^{<>}\left(t'x',tx\right)G_{\text{L}}^{><}\left(tx,t'x'\right)e^{i2k_{F}\left(r-r'\right)}\right]\nonumber 
\end{eqnarray}
with $x=ra_{0}$ , $x'=r'a_{0}$ and $G_{l}\left(zx,z'x'\right)=-i\av{T_{\gamma}\psi_{l}(zx)\psi_{l}^{\dagger}(zx)}$,
since the cross terms $\av{T_{\gamma}\psi_{\text{R}}(zx)\psi_{\text{L}}^{\dagger}(zx)}$
vanish. Defining 
\begin{eqnarray}
G_{l}^{a}\left(tx,t'x'\right) & = & \int\frac{d\omega}{2\pi}\int\frac{dq}{2\pi}e^{-i\omega\left(t-t'\right)}e^{iq\left(x-x'\right)}G_{l}^{a}\left(\omega k\right)
\end{eqnarray}
we have that, for the quantities $\chi_{p}^{\pm}\left(\nu\right)=-\frac{1}{2\pi i}\left[\chi_{p}^{>}\left(\nu\right)\pm\chi_{p}^{<}\left(\nu\right)\right]$,
\begin{eqnarray}
\chi_{p}^{\pm}\left(\nu\right) & = & \left[\chi_{\text{LL}}^{\pm}\left(\nu,a_{0}^{-1}p\right)+\chi_{\text{RR}}^{\pm}\left(\nu,a_{0}^{-1}p\right)\right]+\chi_{\text{LR}}^{\pm}\left[\nu,a_{0}^{-1}\left(p-2k_{F}\right)\right]+\chi_{\text{RL}}^{\pm}\left[\nu,a_{0}^{-1}\left(p+2k_{F}\right)\right]
\end{eqnarray}
with
\begin{eqnarray}
\chi_{ll'}^{\pm}\left(\nu q\right) & = & -\frac{1}{2\pi i}\left[\chi_{ll'}^{>}\left(\nu q\right)\pm\chi_{ll'}^{<}\left(\nu q\right)\right]\nonumber \\
 & = & -\pi\int\frac{d\omega}{2\pi}\int\frac{dk}{2\pi}\left[\rho_{l}^{\pm}\left(\omega k\right)\rho_{l'}^{+}\left(\omega-\nu;k-q\right)-\rho_{l}^{\mp}\left(\omega k\right)\rho_{l'}^{-}\left(\omega-\nu;k-q\right)\right].
\end{eqnarray}
Explicitly, we obtain 
\begin{eqnarray}
\chi_{\text{LL}}^{+}\left(\nu q\right) & = & -\frac{1}{2v}\delta\left(\nu-vq\right)\int\frac{d\omega}{2\pi}\left[F_{\text{L}}\left(\omega\right)F_{\text{L}}\left(\omega-\nu\right)-1\right],\\
\chi_{\text{RR}}^{+}\left(\nu q\right) & = & -\frac{1}{2v}\delta\left(\nu+vq\right)\int\frac{d\omega}{2\pi}\left[F_{\text{R}}\left(\omega\right)F_{\text{R}}\left(\omega-\nu\right)-1\right],\\
\chi_{\text{LR}}^{+}\left(\nu q\right) & = & -\frac{1}{2v}\frac{1}{2\pi}\left[F_{\text{L}}\left(\frac{vq+\nu}{2}\right)F_{\text{R}}\left(\frac{vq-\nu}{2}\right)-1\right],\\
\chi_{\text{RL}}^{+}\left(\nu q\right) & = & -\frac{1}{2v}\frac{1}{2\pi}\left[F_{\text{R}}\left(\frac{-vq+\nu}{2}\right)F_{\text{L}}\left(\frac{-vq-\nu}{2}\right)-1\right],
\end{eqnarray}
and 
\begin{eqnarray}
\chi_{\text{LL}}^{-}\left(\nu q\right) & = & -\frac{1}{2v}\delta\left(\nu-vq\right)\int\frac{d\omega}{2\pi}\left[F_{\text{L}}\left(\omega-\nu\right)-F_{\text{L}}\left(\omega\right)\right],\\
\chi_{\text{RR}}^{-}\left(\nu q\right) & = & -\frac{1}{2v}\delta\left(\nu+vq\right)\int\frac{d\omega}{2\pi}\left[F_{\text{R}}\left(\omega-\nu\right)-F_{\text{R}}\left(\omega\right)\right],\\
\chi_{\text{LR}}^{-}\left(\nu q\right) & = & -\frac{1}{2v}\frac{1}{2\pi}\left[F_{\text{R}}\left(\frac{vq-\nu}{2}\right)-F_{\text{L}}\left(\frac{vq+\nu}{2}\right)\right],\\
\chi_{\text{RL}}^{-}\left(\nu q\right) & = & -\frac{1}{2v}\frac{1}{2\pi}\left[F_{\text{L}}\left(\frac{-vq-\nu}{2}\right)-F_{\text{R}}\left(\frac{-vq+\nu}{2}\right)\right].
\end{eqnarray}

\end{widetext}
\end{document}